\documentclass[reprint,amsmath,amssymb,aps,prl,
floatfix,]{revtex4-2}

\usepackage{graphicx}
\usepackage{dcolumn}
\usepackage{bm}

\usepackage{tikz}
\usetikzlibrary {patterns}
\usepackage{newfloat,algcompatible}
\usepackage[size=small]{caption}
\usepackage{etoolbox}
\usepackage{multirow}
\usepackage{ctable}

\AtBeginEnvironment{algorithm}{\noindent\hrulefill\par\nobreak\vskip-5pt}
\DeclareFloatingEnvironment[
fileext=loa,
listname=List of Algorithms,
name=ALGORITHM,
placement=tbhp,
]{algorithm}
\DeclareCaptionFormat{algorithms}{\vskip-15pt\hrulefill\par#1#2#3\vskip-6pt\hrulefill}
\algnewcommand\algorithmicreturn{\textbf{return}}
\algnewcommand\RETURN{\State \algorithmicreturn}%

\usepackage{newfloat}
\renewcommand{\thefigure}{S\arabic{figure}}

\newtheorem{theorem}{Theorem}
\newtheorem{remark}{Remark}

\begin{document}

\preprint{APS/123-QED}

\title{Upper limits on the robustness of Turing models \\ and other multiparametric dynamical systems}

\author{Roozbeh H. Pazuki}
\author{Robert G. Endres}%
\affiliation{Department of Life Sciences, Imperial College, London SW7 2AZ, United Kingdom}
\affiliation{Centre for Integrative Systems Biology and Bioinformatics, Imperial College, London SW7 2AZ, United Kingdom}

\date{\today}

\begin{abstract}
Traditional linear stability analysis based on matrix diagonalization is a computationally intensive $O(n^3)$ process for $n$-dimensional systems of differential equations, posing substantial limitations for the exploration of Turing systems of pattern formation where an additional wave-number parameter needs to be investigated. In this study, we introduce an efficient $O(n)$ technique that leverages Gershgorin’s theorem to determine upper limits on regions of parameter space and the wave number beyond which Turing instabilities cannot occur. This method offers a streamlined avenue for exploring the phase diagrams of other complex multiparametric models, such as those found in systems biology.
\end{abstract}

\maketitle

Deciphering reproducible pattern formation during embryogenesis with multiparameter models calls for mechanisms demonstrating robustness in parameter space. Turing models of diffusion-driven instability \cite{turing_1952}, which serve as a pivotal category of models, only induce patterns within a highly restricted region of the parameter space \cite{maini_2012,scholes_2019, Hass2021, woolley_2021}. The exploration of such models for increased regions of parameter space and hence enhanced robustness is markedly limited due to their inherent complexity: they incorporate numerous parameters, including rate constants and diffusion constants of activator and inhibitor morphogens, as well as a wave number parameter as part of the analysis. 

Here, we consider autonomous spatiotemporal dynamic models that include diffusion, such as the ones describing reaction-diffusion phenomena. Their partial differential equations (PDEs) are  written as  
\begin{equation}
    \frac{d \bm{X}}{dt}  =  \bm{D} \nabla^2 \bm{X} +  \bm{f}(\bm{X}; \bm{\theta}),\label{eq:RDmodel}
\end{equation}
where $\bm{X} \in \mathbb{R}^n$ are system variables, $\bm{f}$ is an $n$-valued function defined in $n$-dimensional phase space, $\bm{\theta} \in \mathbb{R}^m$ is the system-independent parameter vector, $\bm{D}$ is the diffusion matrix, and $\nabla^2$ the Laplacian.

Generally, when analyzing a  dynamical system, after finding all the fixed points, one needs to study their stability and, eventually, the basins of attraction of all fixed points in the phase space. To elaborate on this procedure, for systems with diffusion-induced instability capable of producing Turing patterns, one requires the system without diffusion to be stable. Hence, one needs to solve the system of equations 
\begin{equation}
    \bm{f}(\bm{X}^*; \bm{\theta}) = 0
\end{equation}
for a given $\bm{\theta}$ to find fixed points $\bm{X}^*$, and then use the Taylor expansion of the equations around these points to linearise the model to study the stability of the system due to a small perturbation for $t \ll 1$. This procedure is equivalent to writing the $n\times n$ Jacobian matrix $\left. \boldsymbol{J}_0 \right|_{\boldsymbol{X}^*}$ of the system at $\bm{X}^*$ and studying its eigenvalues to classify the fixed point stability.
However, finding the eigenvalues of a matrix is computationally expensive $O(n^3)$, so the above procedure is challenging for high-dimensional phase spaces.
The mentioned problem becomes more acute when considering that a candidate dynamical system can have some parameters that change the system's stability. In this case, we need to redo all the steps for every point in parameter space to study the dynamical system's characteristics, e.g. the bifurcation diagram.

Specifically, for our reaction-diffusion system with diffusion-induced instability, the Laplacian is removed by spatial Fourier transformation at the expense of an extra parameter, the wave number. Hence, the linear stability analysis must be applied for different wave numbers to specify the dominant wavelength that finally shapes the stationary solution \cite{scholes_2019}. For a diagonal diffusion matrix $\bm{D}$, the Jacobian for a given wave number, say $k$, is
\begin{equation} \label{eq:jacobian_with_diff}
\begin{split}
 \boldsymbol{J}(k) & = \left. \boldsymbol{J}_0 \right|_{\boldsymbol{X}^*} - k^2 \bm{D} \\
 & = \begin{pmatrix}
        \partial_1 f_1 &  \dots & \partial_n f_1 \\
        \vdots & \ddots & \vdots \\
        \partial_1 f_n & \dots & \partial_n f_n \\
    \end{pmatrix} - k^2
    \begin{pmatrix}
        D_1 & \dots & 0 \\
        \vdots & \ddots  & \vdots \\
        0 & \dots & D_n \\
    \end{pmatrix}.
\end{split}
\end{equation}
Hence, any improvement that decreases the procedure's computational complexity is beneficial for general stability analysis, particularly for pattern formation study. 

In this paper, we introduce an efficient $O(n)$ methodology based on the well-known Gershgorin's theorem \cite{bell_gershgorins_1965} in linear algebra. Our method serves as an initial screening process to rule out extensive portions of the parameter space when searching for Turing instabilities. This method, applicable numerically and analytically for simplified models, markedly accelerates the exploration process. We illustrate the efficacy of this approach using three 2-morphogen models – a sigmoidal Hill-function-based model, the Brusselator model \cite{Prigogine1985}, and the Lengyel-Epstein \cite{Lengyelb1990} model.

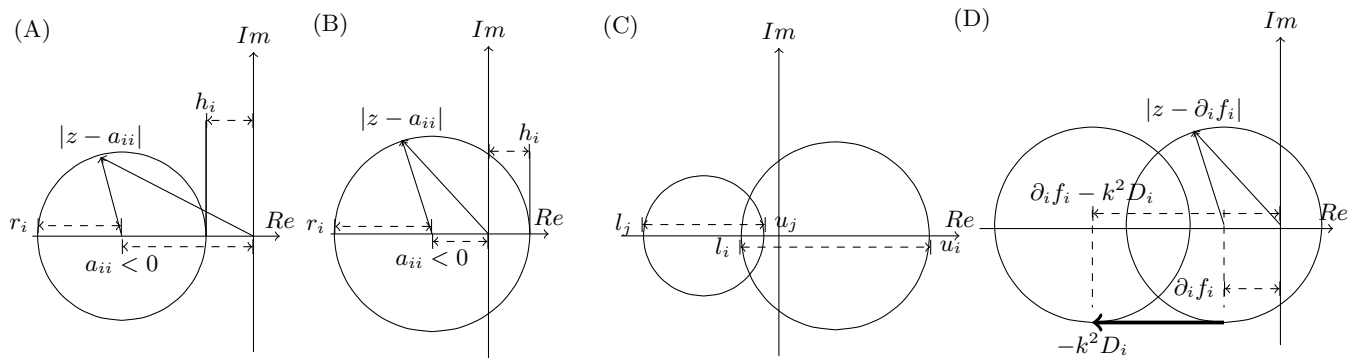
\begin{figure*}[t!]
        \begin{minipage}{.18\textwidth} 
           \begin{tikzpicture}[scale=0.7]
              \draw[] (-4.2,3.5) node[above] {(A)};
              \draw[->] (-4.2,0) -- (0.5,0) node[above] {$Re$};
              \draw[->] (0,-2.2) -- (0,3.50) node[above] {$Im$};
            
              \draw[->] (0.0,0.0) -- (-2.9,1.5) node[above] {$|z - a_{ii}|$};
              \draw[->] (-2.5,0.0) -- (-2.9,1.5);
              \draw [|<->|, dashed] (0,-0.2) --  (-2.5,-0.2) node[below] {$a_{ii} < 0 $};
              \draw [|<->|, dashed] (0,2.2) --  (-0.9,2.2) node[above] {$h_i$};
              \draw [-] (-0.9,0.0) --  (-0.9,2.2);
              \draw [|<->|, dashed] (-2.5,0.2) --  (-4.1,0.2) node[left] {$r_i$};
              \draw [-] (-0.9,0.0) --  (-0.9,2.2);
              \draw[scale=1.0,domain=-3.141:3.141,smooth,variable=\t]
              plot ({-2.5+1.6*sin(\t r)},{1.6*cos(\t r)});
            \end{tikzpicture}            
        \end{minipage} \hfill
        \begin{minipage}{.18\textwidth}
           \begin{tikzpicture}[scale=0.5]
           \draw[] (-4.2,5.0) node[above] {(B)};
          \draw[->] (-4.2,0) -- (1.6,0) node[above] {$Re$};
          \draw[->] (0,-3.3) -- (0,5.0) node[above] {$Im$};
        
          \draw[->] (0.0,0.0) -- (-2.3,2.5) node[above] {$|z - a_{ii}|$};
          \draw[->] (-1.5,0.0) -- (-2.3,2.5);
          \draw [|<->|, dashed] (0,-0.2) --  (-1.5,-0.2) node[below] {$a_{ii} < 0$};
          \draw [|<->|, dashed] (0,2.2) --  (1.1,2.2) node[above] {$h_i$};
          \draw [|<->|, dashed] (-1.5,0.2) --  (-4.1,0.2) node[left] {$r_i$};
          \draw [-] (1.1,0.0) --  (1.1,2.2);
          \draw[scale=1.0,domain=-3.141:3.141,smooth,variable=\t]
          plot ({-1.5+2.6*sin(\t r)},{2.6*cos(\t r)});
        \end{tikzpicture}
        \end{minipage} \hfill
        \begin{minipage}{.22\textwidth}
          \begin{tikzpicture}[scale=.5,domain=0:4]
          \draw[] (-4.2,5.0) node[above] {(C)};
          \draw[->] (-4.2,0) -- (4.8,0) node[above] {$Re$};
          \draw[->] (0,-3.2) -- (0,5.0) node[above] {$Im$};
          \draw[scale=1.0,domain=-3.141:3.141,smooth,variable=\t]
          plot ({1.5+2.5*sin(\t r)},{2.5*cos(\t r)});
          \draw[scale=1.0,domain=-3.141:3.141,smooth,variable=\t]
          plot ({-2.0+1.6*sin(\t r)},{1.6*cos(\t r)});
          
          \draw [|<->|, dashed] (-1.05,-0.3) node[left] {$l_i$} --  (4.05,-0.3) node[right] {$u_i$};

          \draw [|<->|, dashed] (-0.35,0.3) node[right] {$u_j$} --  (-3.65,0.3) node[left] {$l_j$};      
         \end{tikzpicture}
        \end{minipage} \hfill   
        \begin{minipage}{.26\textwidth}
        \begin{tikzpicture}[scale=.5]
          \draw[] (-8.2,4.5) node[above] {(D)};
          \draw[->] (-8.0,-.50) -- (1.4,-.50) node[above] {$Re$};
          \draw[->] (0,-3.6) -- (0,4.6) node[above] {$Im$};
        
          \draw[->] (0.0,-0.4) -- (-2.3,2.1) node[above] {$|z - \partial_i f_i|$};
          \draw[->] (-1.5,-0.4) -- (-2.3,2.1);
          \draw [|<->|, dashed] (0,-2.1) --  (-1.5,-2.1) node[left] {$\partial_i f_i$};   
          \draw[scale=1.0,domain=-3.141:3.141,smooth,variable=\t]
          plot ({-1.5+2.6*sin(\t r)},{2.6*cos(\t r)-.4});

          \draw[scale=1.0,domain=-3.141:3.141,smooth,variable=\t]
          plot ({-5.0+2.6*sin(\t r)},{2.6*cos(\t r)-.4});
          \draw [|<->|, dashed] (0,-0.1) --  (-5.0,-0.1) node[above] {$\partial_i f_i -k^2 D_i$}; 

          \draw [->, line width=.5 mm] (-1.5,-3.0) --  (-5.0,-3.0) node[below] { $-k^2 D_i$}; 
          \draw [dashed] (-5.0,0) --  (-5.0,-2.6);
          \draw [dashed] (-1.5,0) --  (-1.5,-2.6);
        \end{tikzpicture}
        \end{minipage} \hspace*{\fill}   
   \caption{\textbf{Gershgorin's theorem and its geometrical interpretation.} (A) Type 1 (stable). (B) Type 2 (inconclusive). (C) Two overlapping circles. (D) Diagonal terms shift to the left when diffusion is included.}
    \label{fig:1}
\end{figure*}

\textbf{Gershgorin's theorem} -
We start by introducing Gershgorin's theorem and consider its geometrical interpretation. As we shall see, it is possible to construct an algorithm that checks the rows or columns of a Jacobian matrix to find those that are unstable or remain stable after introducing diffusion and consequently cannot produce a Turing pattern. The theorem \cite{bell_gershgorins_1965} states that for $n \times n$ complex matrix $A=(a_{ij})$ and $r_i \equiv \sum_{\substack{j=1 \\ j \ne i}}^{n} |a_{ij}|$ the sum of moduli of off-diagonal elements in the $i$-th row, each union of circles $|z - a_{ii}| \le  r_i$ (for $i = 1, 2, \dots, n$) contains a number of eigenvalues of $A$ equal to the number of circles used to create the union.
The analogous result holds if columns of $A$ are considered. Note that in our case, $a_{ij} = \partial_i f_j$, and for diagonal terms $a_{ii} = \partial_i f_i$ without and $a_{ii} = \partial_i f_i - k^2 D_i$ with diffusion.

Let us consider the radius and position of a circle in the complex plane as depicted in Fig. \ref{fig:1}. Depending on the sign of the diagonal term $a_{ii}$, the circle's center is on the real axis's negative or positive side. Let us say $\lambda_i$ is one of the eigenvalues corresponding to the row or column $i$ of the matrix $A$. Four types of circles may appear: {\it Type 1}: As it shown in Fig. \ref{fig:1}A, we have $ a_{ii} < 0$ and $h_i \equiv |a_{ii}| - r_i \ge 0$.
Therefore, regardless of where the eigenvalue is inside the circle, its real part must be negative, $Re(\lambda_i) \le 0$. {\it Type 2}: In Fig. \ref{fig:1}B, we can see $a_{ii} <0$, and the center of the circle is placed on the negative side of the real axis. However, since $|a_{ii}| < r_i$ and $h_i < 0$, the real part of $\lambda_i$ can be negative or positive. In other words, the theorem is inconclusive about the sign of the real part of the corresponding eigenvalue. {\it Type 3}: The diagonal element is positive ($a_{ii} > 0$), and the center of the circle is on the positive side of the real axis. Yet, $|a_{ii}| > r_i$ and $h_i > 0$. Thus, the real part of the eigenvalue must be positive $Re(\lambda_i) > 0$.{\it Type 4}: Similar to Type 2, the range of Gershgorin's circle spans from negative to positive values. Therefore we cannot conclusively decide about the real part of the eigenvalue inside this area.

We must emphasize that when two or more Gershgorin's circles overlap, the eigenvalues lie in the union of the circles. Let us define the lower and upper bounds of each circle as $l_i \equiv a_{ii} - r_i$ and $u_i \equiv a_{ii} + r_i$, respectively. For instance, in Fig. \ref{fig:1}C, the union of two circles shows that the real part of both eigenvalues corresponding to rows $i$ and $j$ must be in the union of their diameters $Re(\lambda_i), Re(\lambda_j) \in [l_j, u_j] \bigcup [l_i, u_i] = [l_j, u_i]$.  And correspondingly, for each row (or column) of the matrix $A$, there exists intervals like 
\begin{equation}
       A= \begin{pmatrix}
        a_{11} & a_{12} & \dots & a_{1n}\\
        a_{21} & a_{22} & \dots & a_{2n}\\
        \vdots & \quad  &\ddots & \vdots \\
        a_{n1} & a_{n2} & \dots & a_{nn}\\
        \end{pmatrix}
        \implies
        \begin{matrix}
        [l_1, u_1] \\
        [l_2, u_2] \\
        \vdots \\
        [l_n, u_n] \\
        \end{matrix},
\end{equation}
where after taking their unions and reformulating as non-overlapping, disjoint intervals result in 
\begin{equation}
    \mathcal{A} = \bigcup_{i=1}^{n} [l_i, u_i] \equiv \bigcup_{i=1}^{p}  [L_i, U_i] , \qquad 1 \le p  \le n,
\end{equation}
such that $[L_i, U_i] \cup [L_j, U_j] = \emptyset$, for $i \ne j$. Therefore, the $[L_{max}, U_{max}]$ is the right-most disjoint interval constructed by the unions of the original intervals, and is sufficient to study it to find the sign of the real part of the largest eigenvalue. This introduces the following possibilities regarding the stability condition: 1. For $U_{max} \le 0$, the real part of the largest eigenvalue is negative. Consequently, the system is stable. 2. For $L_{max} > 0$, the real part of the largest eigenvalue is positive. Consequently, the system is unstable. 3. For $L_{max} \le 0 < U_{max}$ the situation is inconclusive.

At this stage, we can use the obtained results in two different ways. The first possibility is when the Jacobian is written in terms of the model's parameters, $\bm{\theta}$, and we may be able to derive the right-most disjoint set parametrically. Accordingly, our inequalities define the regions in parameter space where the method can conclusively determine the stability/instability of the system. Nevertheless, the region corresponding to the inconclusive range requires different classification techniques. Note that even if the method is not always conclusive for all the regions in parameter space, it can always find a theoretical lower bound for the volume of the parameter space that the system is stable/unstable. 

The second possibility is when studying a system's linear stability numerically. We propose an algorithm  that classifies a given Jacobian matrix into ``\textit{stable}'', ``\textit{unstable}'' and ``\textit{inconclusive}'' stability groups (see Algorithm (1) in Supplementary Materials - IV A).

\textbf{Reaction-diffusion models} - 
To study a pattern-forming system given by Eq. (\ref{eq:RDmodel}) with stationary solution $\bm{X}^* = (X^*_1, \dots, X^*_n)$, the linear stability of the Jacobian $\left. \bm{J} \right|_{\bm{X}^*}$ in Eq. (\ref{eq:jacobian_with_diff}) is studied by our proposed method. We can write the lower and upper bounds corresponding to row (or column) $i$ as $l_i = \partial_i f_i - r_i$ and $u_i = \partial_i f_i + r_i$ for $r_i \equiv \sum_{j \ne i} \partial_j f_i$.
Then, the stability/instability criteria of the Jacobian determine the Turing pattern conditions. 

It is interesting to see the effect of introducing diffusion. 
For a given wave number $k$, the inclusion of diffusion shifts all the diagonal terms by $-k^2 D_i$ (see Eq. (\ref{eq:jacobian_with_diff})).
Effectively, since diagonal terms correspond to the location of the centers of Gershgorin's circles, it is geometrically equivalent to saying all the circles shift to the left as shown in Fig. \ref{fig:1}D. Note, since the off-diagonal terms have not changed, the circles' radii remain unchanged.

Indeed, for any given circle partially on the positive side of the real axis, there exists a maximum shift by a wave number defined by $k^{*}_i=\sqrt{(r_i + \partial_i f_i)/D_i}$ that transfers the circle entirely to the negative side of the real axis by $k^{*2}_i D_i$ (for details see Supplementary Materials - II). Furthermore, the real part of the eigenvalue corresponding to that circle must be negative for all the higher wave numbers $k_i > k^{*}_i$. By finding $k^{*}_{max} \equiv \max \{k^{*}_{1}, \dots, k^{*}_{n} \}$ for all rows (or columns), the linear stability of different wave numbers can be restricted to the range $k$, such that for $k=0$, or the case with no diffusion, $\left. \boldsymbol{J}_0 \right|_{\boldsymbol{X}^*}$ specifies the stability condition and $\boldsymbol{J}(k)$ for $k \in (0, k^{*}_{max}]$ determines the evolution of the dominant wave number in a perturbed system.

Thus, we must have three different regimes: (1) For $\left. \boldsymbol{J}_0 \right|_{\boldsymbol{X}^*}$, when $U_{max} \le 0$, all the circles are on the negative side of the real line and including diffusion for any given wave number shifts the circles further to the left. Hence, all real parts of eigenvalues are negative, and diffusion cannot excite any wave number. Consequently, the system is incapable of producing a Turing pattern (``\textit{super-stable}'').
(2) On the contrary, when $L_{max} > 0$ for $\left. \bm{J}_0 \right|_{\bm{X}^*}$, the initial stationary state is unstable and incapable of producing a Turing pattern (``\textit{unstable}''). (3) Finally, when $[L_{max}, U_{max}]$ is inconclusive ($L_{max} <0 $ and $U_{max} > 0$), $\left. \boldsymbol{J}_0 \right|_{\boldsymbol{X}^*}$ must be studied by finding its eigenvalues. And if one finds that it is stable, the maximum of the dispersion relation $\lambda(k)$ finds the dominant wave number for pattern formation by restricting $k$ to $[0, k^{*}_{max}]$. These regimes are included in our algorithm to speed up the process for checking the possibility of Turing-pattern formation for a given parameter set. Only parameters for which the classification is ``inconclusive" need further study of their eigenvalues and are in principle able to form patterns.

\textbf{Role of diffusing species} -
The dispersion relation of systems in which all species are diffusing always satisfies $Re(\lambda(k)) < 0$ for $k > k^{*}_{max}$. In other words, asymptotically, as long as all species are diffusers, we have $\lim_{k \rightarrow \infty} Re(\lambda(k)) \rightarrow -\infty$. When some but not all species in a reaction-diffusion model diffuse, introducing the diffusion coefficients into the Jacobian matrix shifts some circles to the left while the others remain in the same place -- see Fig. \ref{fig:2}. Although one can calculate the $k^{*}_{max}$ values for the diffuser rows in the matrix, a special situation can arise for $k > k^{*}_{max}$ in the dispersion relation. For instance, consider two among three species are diffusers as shown in Fig. \ref{fig:2}. As we can see, after shifting the diffusers circle by an amount $-k^{*2}_{max} D_1$ and $-k^{*2}_{max} D_2$ respectively, the third eigenvalue corresponds to the non-diffuser element remains positive for all $k > k^{*}_{max}$. Consequently, the real part of the dispersion relation can remain positive with no upper bounds. As a result, no dominant wave number exists to create a stationary pattern. Fortunately, it is easy to find these cases algorithmically without calculating the eigenvalues.

\begin{figure}
    \begin{minipage}{.5\textwidth}
        \begin{tikzpicture}[scale=.6]
        \draw  (-7.7,4.0) node[right] {(A)};          
        
          \draw[->] (-5.8,0) -- (4.0,0) node[right] {$Re$};
          \draw[->] (0,-2.8) -- (0,3.0) node[above] {$Im$};

         \draw[-, dashed]  (-1.5,0.8) --  (-1.5, 0.0);
          \draw [|<->|, dashed] (0,0.8) --  (-1.5,0.8) node[left] {$\partial_1 f_1$};          
          \draw[scale=1.0,domain=-3.141:3.141,smooth,variable=\t]
          plot ({-1.5+2.6*sin(\t r)},{2.6*cos(\t r)});
        \draw[-, dashed]  (1.0,-2.2) --  (1.0, 0.0); 
        \draw [|<->|, dashed] (0,-2.2) --  (1.0,-2.2) node[right] {$\partial_2 f_2$};   \draw[scale=1.0,domain=-3.141:3.141,smooth,variable=\t]
          plot ({1.0+1.6*sin(\t r)},{1.6*cos(\t r)});
          \draw[-, dashed]  (2.5,2.2) --  (2.5, 0.0);
          \draw [|<->|, dashed] (0,2.2) --  (2.5,2.2) node[right] {$\partial_3 f_3$};   \draw[scale=1.0,domain=-3.141:3.141,smooth,variable=\t]
          plot ({2.5+.95*sin(\t r)},{.95*cos(\t r)});               
        \end{tikzpicture}
    \end{minipage} \hfill
    \begin{minipage}{.5\textwidth} 
         \begin{tikzpicture}[scale=.6]
          \draw  (-7.5,4.0) node[right] {(B)};
        
          \draw[->] (-5.5,0) -- (4.0,0) node[right] {$Re$};
          \draw[->] (0,-3.2) -- (0,3.5) node[above] {$Im$};

          \draw[-, dashed]  (-2.6,3.2) --  (-2.6, 0.0);
          \draw [|<->|, dashed] (0,3.2) --  (-2.6,3.2) node[left] {$\partial_1 f_1 - k^{*2}_{max} D_1$};          
          \draw[scale=1.0,domain=-3.141:3.141,smooth,variable=\t]
          plot ({-2.6+2.6*sin(\t r)},{2.6*cos(\t r)});

          \draw[-, dashed]  (-2.1,-2.4) --  (-2.1, 0.0);
        \draw [|<->|, dashed] (0,-2.4) --  (-2.1,-2.4) node[below] {$\partial_2 f_2 - k^{*2}_{max} D_2$};   \draw[scale=1.0,domain=-3.141:3.141,smooth,variable=\t]
          plot ({-2.1+1.6*sin(\t r)},{1.6*cos(\t r)});

          \draw[-, dashed]  (2.5,2.2) --  (2.5, 0.0);
          \draw [|<->|, dashed] (0,2.2) --  (2.5,2.2) node[right] {$\partial_3 f_3$};    \draw[scale=1.0,domain=-3.141:3.141,smooth,variable=\t]
          plot ({2.5+.95*sin(\t r)},{.95*cos(\t r)});
               
        \end{tikzpicture}
    \end{minipage}\hspace*{\fill}
    \caption{{\bf Maximum shift of two diffusers.} (A) Circles before inclusion of diffusion. (B) The maximum shifts of the first and second circles due to diffusion.}
    \label{fig:2}
\end{figure}
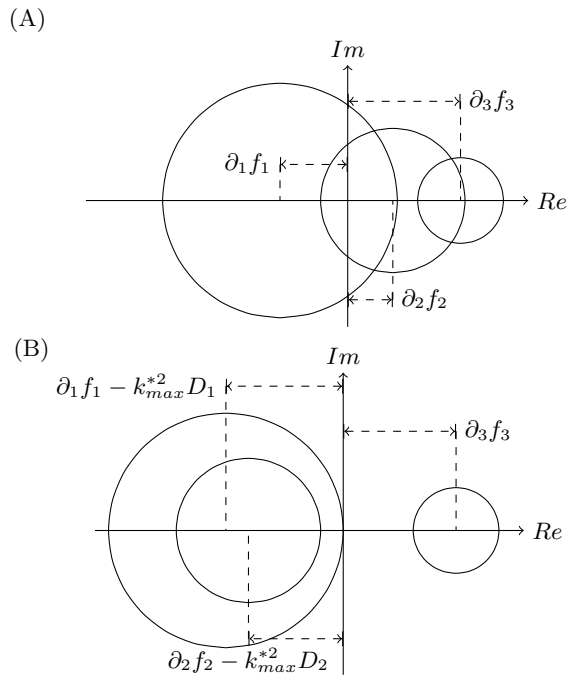

\begin{figure}
    \begin{minipage}{.5\textwidth}
    \begin{tikzpicture}[scale=.6]
            \draw[] (-2, 6.0) node[above] {(A)};
		\draw[->] (-.2,0) -- (7.5,0) node[right] {$a$};
		\draw[->] (0,-0.75) -- (0,5.5) node[above] {$b$};
      \draw[domain=0:2.0,smooth,variable=\a]
		plot ({\a},{1 + \a * \a}) node[right] {$b = 1 + a^2$}; 
		\fill[pattern=crosshatch,domain=0:2.0,smooth,variable=\a]
		plot ({\a},{1 + \a * \a}); 
		\fill[pattern=crosshatch] (2.0, 5.0) -- (0.0, 5.0) -- (0.0, 1.0);          
		\draw[domain=0:1.2,smooth,variable=\a]
		plot ({\a},{1 - \a * \a}) node[right] {$b = 1 - a^2$}; 
		\filldraw[pattern=crosshatch,domain=0.707106:1.0,smooth,variable=\a]
		plot ({\a},{1 - \a * \a})  -- (0,0) -- (0.707106, 0); 
		
		\draw[domain=0:2.0,smooth,variable=\a]
		plot ({\a},{ \a * \a}) node[right] {$b = a^2$};
		\filldraw[pattern=crosshatch,line width=0pt,domain=0:0.707106,smooth,variable=\a]
		plot ({\a},{ \a * \a}) -- (0.707106, 0) -- (0,0); 
		
		\draw[domain=0:7.0,smooth,variable=\a]
		plot ({\a},{ .5}) node[right] {$b = 1/2$};
		
		\draw[domain=0:7.0,smooth,variable=\a]
		plot ({\a},{ 1}) node[right] {$b = 1$};
		
		\end{tikzpicture}
    \end{minipage} 
    \begin{minipage}{.5\textwidth}
    \begin{tikzpicture}[scale=.6]
    \draw[] (-2.0, 4.0) node[above] {(B)};
		\draw[->] (-.2,0) -- (7.5,0) node[right] {$a$};
		\draw[->] (0,-0.75) -- (0,3.5) node[above] {$b$};
		\draw[domain=0:7.0,smooth,variable=\a]
		plot ({\a},{ 1}) node[right] {$b = 1$};
		
		\draw[domain=0:7.0,smooth,variable=\a]
		plot ({\a},{ .5}) node[right] {$b = 1/2$};
		
		\filldraw[pattern=crosshatch,domain=0:7.0,smooth,variable=\A]
		plot ({\A},{ .5}) -- (7.0, 0.0) -- (0,0);
		
		\end{tikzpicture}
    \end{minipage}    
    \caption{\textbf{Brusselator's parameter space}. (A) Row-induced parts of the parameter space (hatched area) that cannot produce Turing patterns. (B) Column-induced conclusive part of the parameter space (hatched area).}
    \label{fig:3}
\end{figure}

\textbf{Tightening the bounds} -
Given an invertible matrix $\bm{D}$, $\bm{B} = \bm{D} \bm{A} \bm{D}^{-1}$ introduces an equivalence relation between square matrices $\bm{A}$ and $\bm{B}$ such that matrix $\bm{B}$ has the same eigenvalues as $\bm{A}$ (see Supplementary Material - III for details).
Defining an invertible $n \times n$ diagonal matrix $\bm{D}$ as the identity matrix with exception of matrix element $D_{ii}=1/d_i$,
the transformed matrix $\bm{D} A \bm{D}^{-1}$ has the form
\begin{equation}
 \bm{D} \bm{A} \bm{D}^{-1} = \begin{pmatrix}
        a_{11}  & \dots & a_{1i} d_i &\dots & a_{1n} \\        
        \vdots & \ddots & \vdots & \quad & \vdots \\       
        \frac{a_{i1}}{d_i}  & \dots & a_{ii} &\dots & \frac{a_{in}}{d_i} \\
        \vdots & \quad & \vdots & \ddots &  \vdots \\
        a_{n1} & \dots & a_{ni} d_i &\dots &  a_{nn} \\
    \end{pmatrix}.    
\end{equation}
The transformation's effect is similar to dividing all the elements of the row $i$ by $d_i$ and multiplying the elements of the column $i$ by $d_i$. Consequently, the diagonal term $a_{ii}$ remains the same.

Consider the rows of the resulting matrix. The radii of all circles corresponding to rows other than $i$ expand by the amount $|a_{ji}| (d_i-1)$ (for $d_i > 1$), and the $i$-th radius shrinks by the factor of $1/d_i$, while the centers of all circles stay the same. Since the eigenvalues of the transformed matrix are the same as the original one, one can hope the shrunk circle becomes isolated from the rest since the expansions of the other radii are smaller than the shrinking of the single radius. In practice, one can find $d_i$ that isolates the circle with the largest center from the rest.
The interval of all rows except the $i$-th is $[l_j, u_j]=\left[a_{jj} - |a_{ji}| (d_i - 1)  -  r_j , \quad a_{jj} + |a_{ji}| (d_i-1) +  r_j \right]$, and the interval for row $i$ is $[l_i, u_i] = \left[a_{ii} -  \frac{r_i}{d_i} , \quad a_{ii}  +  \frac{r_i}{d_i} \right]$.

To tighten the bounds, we have two distinct cases that can be studied separately. Case (1): When the diagonal term is positive, or $a_{ii} > 0$, to isolate the circle corresponding to row $i$, its leftmost point, or $l_i = a_{ii} -  \frac{r_i}{d_i}$, must be larger than every other circle's rightmost point, or $u_j =  a_{jj} + |a_{ji}| (d_i-1) +  r_j$. As explained in detail in Supplementary Materials - III, if all the rows $j$ and the largest one, $i$, satisfy the inequalities
\begin{eqnarray}
    \begin{cases}        
    \begin{matrix}
      (a_{jj} - a_{ii} +  r_j - |a_{ji}|)^2 &>& 4 |a_{ji}| r_i  & \quad j \ne i \qquad \qquad \quad  \\    
      a_{jj} - a_{ii} +  r_j - |a_{ji}| &<& 0   &   j = 1, \dots, n
    \end{matrix},
    \end{cases}
\end{eqnarray}
simultaneously, there exists a $d_i$ that isolates the rightmost eigenvalue, and the Jacobian is conclusively unstable. Case (2): When the largest diagonal term is negative, or  $a_{ii} < 0$, the latter implies all the other diagonal terms are negative too. In this case, we search for a possible shrinkage value $d_i$, such that while the circle of the row $i$ shrinks with its upper bound on the negative side of the real axis, the growth of all the other circles keeps them at the negative side. Combined, this leads to two conditioned bounds as
 \begin{equation}
     \frac{r_i}{|a_{ii}|} < d_i < \min_{j \ne i} \left(\frac{|a_{jj}| - r_j}{|a_{ji}|} + 1\right).
 \end{equation}
 Hence, if a non-empty interval can be found that satisfies the above inequalities, the Jacobian is conclusively ``super-stable" (see Algorithm 2 in Supplementary Materials - IV B). 

\textbf{Conclusion} - We have introduced an efficient $O(n)$ method that uses Gershgorin’s theorem to define robustness bounds in dynamical systems, particularly beneficial for reaction-diffusion models such as diffusion-driven Turing systems \cite{scholes_2019}. This approach can eliminate unstable non-diffusive cases or solutions that remain stable post-diffusion while also setting an upper limit for wave numbers capable of pattern formation. When applied to specific models in the appendix, our method does not only enhance numerical algorithms' speed, but it also facilitates an analytical study of parameter space. The former capability is exemplified by a Hill-function-based Turing model which leads to a rejection of $99.3\%$ of the parameter combinations in table \ref{table:1}), while the latter is demonstrated for the classic Brusselator model in Fig. \ref{fig:3} (and by the Lengyel-Epstein model in Supplementary Materials - V B). The method’s utility increases significantly when accounting for parameters that alter the behavior of a potential dynamical system. For example, generating bifurcation profiles of dynamical systems necessitates repeated linear stability analysis for many different parameter values, a challenging endeavor in high dimensional phase spaces.

\begin{table*}
\caption{\textbf{Empirical statistics of applying Algorithms (1) and (2) to the Hill-functions-based Turing model.} We selected all $10^9$ parameter combinations from $\{0.01, 0.05, 0.1, 0.5, 1, 5, 10, 50, 100, 500\}^9$, and fixed $D_u=0.01$ and $D_v=1.0$ in Eq. (\ref{eq:Hill-function-pde}). Note that inconclusive cases in each row are used for the next run (``Total'' column), and the percentages are calculated for each row independently. Finally, the ``Combined'' row shows the sums of the columns, except for ``Inconclusive'', which is transferred from the last run. We used the ``hybrd'' implementation in the ``MIPACK'' library for the root-finding  algorithm.}
\label{table:1}
\begin{ruledtabular}
\begin{center}
\begin{tabular}{ cccccc } 
 & \textbf{Total} & \textbf{Super-stable} & \textbf{Inconclusive} & \textbf{Unstable} & \textbf{No fixed point}\\ 
 \hline
 \multirow{2}{6em}{Row-wise} & $10^9$  & $850,677,030$ & $140,394,311$ & $4,870,615$ & $4,058,044$ \rule{0pt}{2.6ex} \\ 
 
 & 100\%  & 85.07\% & 14.04\% & 0.49\% & 0.41\% \rule{0pt}{2.6ex} \\ 
 \hline
 \multirow{2}{6em}{Column-wise}  & $140,394,311$  & $68,454,498$ & $71,913,845$ & $25,968$ & - \rule{0pt}{2.6ex} \\ 
 
 & 100\%  & 48.76\% & 51.22\% & 0.02\% & - \rule{0pt}{2.6ex} \\ 
 \hline
 \multirow{2}{6em}{Tighten bounds}  & $71,913,845$  & $64,615,611$ & $6,990,298$ & $307,936$ & - \rule{0pt}{2.6ex} \\ 
 
 & 100\%  & 89.85\% & 9.72\% & 0.43\% & - \rule{0pt}{2.6ex} \\ 
 \specialrule{.2em}{.1em}{.1em} 
 \multirow{2}{6em}{Combined} & $10^9$  & $983,747,139$ & $6,990,298$ & $5,204,519$ & $4,058,044$ \rule{0pt}{2.6ex} \\ 

 & 100\%  & 98.37\% & 0.70\% & 0.52\% & 0.41\% \rule{0pt}{2.6ex} \\ 
\end{tabular}
\end{center}
\end{ruledtabular}
\end{table*}

\textbf{Appendix on applications to Turing models\label{app}} ---
To test the fraction of rejections and, consequently, the speed up due to the Algorithms (1) and (2), we use a biologically inspired model capable of producing Turing patterns \cite{scholes_2019}. The reaction-diffusion PDE with nine free parameters is written as 
\begin{eqnarray}
\frac{\partial u}{\partial t} &=& b_u  +     \frac{V_u}{\left[1+\left(\frac{K_{uu}}{u }\right)^{4}\right] \left[1+\left(\frac{v}{K_{vu} }\right)^{4}\right]} 
 \nonumber \\
 &&-  \mu_u u + D_u \nabla^2 u, \nonumber\\
 \frac{\partial v}{\partial t} &=& b_v  +     \frac{V_v}{1+\left(\frac{K_{uv}}{u }\right)^{4}}  -  \mu_v v + D_v \nabla^2 v.
 \label{eq:Hill-function-pde}
\end{eqnarray}
Note that the nonlinear terms are Hill functions that regulate activation and inhibition of molecule production, e.g., in gene expression. The Jacobian of the linearized form of the above equations is a two-by-two matrix, and in practice, the computational cost of calculating its eigenvalues is not much different than our algorithm. However, we selected this model since the algorithm's correctness can be easily checked by comparing the determinant and trace of the Jacobian.

In this simulation, we selected one billion parameter combinations and applied Algorithm (1) to classify them into ``\textit{unstable}'', ``\textit{super-stable}'', ``\textit{inconclusive}'' and ``\textit{no fixed point}''. Note that the case ``\textit{no fixed point}'' refers to the parameter combinations for which the root-finding algorithm could not find any stationary solutions. We first used our Algorithm (1) for a row-wise comparison, and after that, by using the inconclusive results from the first run, we used the algorithm again for a column-wise calculation. And finally, we classified the remaining inconclusive cases using our Algorithm (2). These results are presented in table \ref{table:1}, showing that more than $99.3\%$ of the parameter combinations were rejected. This provides an upper limit on the robustness of Turing patterns given a certain sampling of parameter space \cite{scholes_2019}.

\textbf{Brusselator model} -
Next, we look for inequalities that separate the parameter space of the Brusselator model into ``inconclusive`` or otherwise. The Brusselator is a two-species reaction-diffusion model with a set of PDEs given by
\begin{eqnarray}
\frac{d u}{d t} &=& D_u \nabla^2 u + a - (b+1)u + u^2v, \nonumber \\
\frac{d v}{d t} &=& D_v \nabla^2 v + bu - u^2 v,
\end{eqnarray}
for two parameters $a, b > 0$. Using the stability analysis, we can derive the Jacobian of the model at its fixed point $(u^*, v^*) = (a, b/a)$ and the corresponding rows and columns intervals as 
\begin{equation}    
 \begin{matrix}
       \begin{pmatrix}
        b - 1 & \qquad \qquad & a^{2}\\
        - b & \qquad \qquad & - a^{2}
    \end{pmatrix} &
     \begin{matrix}
       \implies & [b-1-a^2, \, b-1+a^2]\\
       \implies & [-a^2-b, \, -a^2+b]
    \end{matrix} \\
    \begin{matrix}
        \Downarrow & \Downarrow \\
        [-1, 2b-1] &
        [-2a^2, 0]
    \end{matrix} & \quad \\
    \end{matrix}.
\end{equation}
Note that since the intervals depend on the parameters, it is easier to check the sign of the circles' centers and classify the conditions than directly taking the union.

For the first row, the circle's center is at $b-1$; therefore, for $0 < b \le 1$, it is on the negative side of the real axis. At the same time, the center of the second row's circle is always on the negative side ($- a^2$). Consequently, the region of parameter space in which the Jacobian is stable before and after the inclusion of diffusion derives as
\begin{equation}
\label{eq:inq-1}
\begin{cases}
    b-1+a^2 \le 0, \\
    -a^2 +b \le 0
    \end{cases} \implies    \begin{cases}
        b \le 1 - a^2, \\
        b \le a^2, \\
    \end{cases},
\end{equation}
for $a, b >0$.
At the same time, for $b > 1$, the center of the first row's circle is on the positive side of the real axis. Thus, the Jacobian is unstable when the first interval's lower bound becomes positive, leading to
\begin{equation}
\label{eq:inq-2}
    \begin{cases}
        b >  a^2 + 1,\\
        a > 0    
    \end{cases}.    
\end{equation}

The conditions from both Eqs. (\ref{eq:inq-1}) and (\ref{eq:inq-2}) are shown in Fig. \ref{fig:3}A.
Similarly, for the column's intervals, if $b < 1/2$, both centers are on the negative sides of the real axis, and the condition of stability writes as $b \le  1/2$ for $a, b > 0$.
However, for $b > 1/2$, the interval of the first column stays between $-1$ and $2b-1$, which means the case is inconclusive. Fig. \ref{fig:3}B depicts the columns' results.

\bibliographystyle{apsrev4-2}
\bibliography{Gershgorin}


\widetext
\newpage
\begin{center}
	\textbf{\large Supplementary Materials:\\Upper limits on the robustness of Turing models \\ and other multiparametric dynamical systems}
\end{center}

\setcounter{equation}{0}
\setcounter{figure}{0}
\setcounter{table}{0}
\setcounter{page}{1}
\makeatletter
\renewcommand{\theequation}{S\arabic{equation}}
\renewcommand{\thefigure}{S\arabic{figure}}
\renewcommand{\bibnumfmt}[1]{[S#1]}
\renewcommand{\citenumfont}[1]{S#1}

In these supplementary materials, we provide details on the results presented in the main text. We also present our Algorithms (1) and (2), and additional results on the Lengyel-Epstein model.

\section{Gershgorin's Theorem}
\begin{theorem}[Gershgorin's Theorem \cite{bell_gershgorins_1965}]
	Let $A=(a_{ij})$ be an $n \times n$ complex matrix, and $\sum_{\substack{j=1 \\ j \ne i}}^{n} |a_{ij}|$ be the sum of moduli of off-diagonal elements in the $i$-th row. Then, each eigenvalue of $A$ lies in the union of the circle
	\begin{equation}
	|z - a_{ii}| \le \sum_{\substack{j=1 \\ j \ne i}}^{n} |a_{ij}| \equiv r_i, \qquad i = 1, 2, \dots, n.
	\end{equation}
	The analogous result holds if columns of $A$ are considered.
\end{theorem}

\subsection{Geometrical interpretation}
Let us consider the radius and position of a circle in the complex plain regarding $i$-th row as depicted in Fig. \ref{fig:1}A, \ref{fig:1}B, \ref{fig:1}C and \ref{fig:1}D. Depending on the sign of the diagonal term $a_{ii}$, the circle's center is either on the real axis's negative or non-negative side. Let us say $\lambda_i$ is one of the eigenvalues corresponding to the row or column $i$ of the matrix $A$. It follows:
\begin{itemize}
	\item \textbf{Type 1}: As it shown in Fig. \ref{fig:1}A, we have $ a_{ii} < 0$, $|a_{ii}| > r_i$ and $h_i = |a_{ii}| - r_i \ge 0$.
	Therefore, regardless of where the eigenvalue is inside the circle, its real part must be negative $Re(\lambda_i) \le 0$.    
	
	\item \textbf{Type 2}: In Fig. \ref{fig:1}B, we can see $a_{ii} <0$, and the center of the circle is placed on the negative side of the real axis. However, since $|a_{ii}| < r_i$ and $h_i < 0$, the real part of $\lambda_i$ can be negative or positive. In other words, the theorem is inconclusive about the sign of the real part of the corresponding eigenvalue.
	
	\item \textbf{Type 3}: The diagonal element is positive ($a_{ii} > 0$), and the center of the circle is on the positive side of the real axis. See Fig. \ref{fig:1}C. Yet, $|a_{ii}| > r_i$ and $h_i > 0$. Thus, the real part of the eigenvalue must be positive $Re(\lambda_i) > 0$.
	\item \textbf{Type 4}: Similar to type 2, the range of Gershgorin's circle spans from negative to positive values. Therefore we cannot conclusively decide about the real part of the eigenvalue inside this area.
\end{itemize}
\begin{figure}
	\begin{minipage}{1\textwidth}
		\begin{minipage}{0.50\textwidth}
			\centering
			\begin{tikzpicture}
			\draw[] (-7, 3.8) node[above] {(A)};
			\draw[->] (-4.2,0) -- (0.5,0) node[right] {$Re$};
			\draw[->] (0,-2.2) -- (0,3.50) node[above] {$Im$};
			
			\draw[->] (0.0,0.0) -- (-2.9,1.5) node[above] {$|z - a_{ii}|$};
			\draw[->] (-2.5,0.0) -- (-2.9,1.5);
			\draw [|<->|, dashed] (0,-0.2) --  (-2.5,-0.2) node[below] {$a_{ii} < 0 $};
			\draw [|<->|, dashed] (0,2.2) --  (-0.9,2.2) node[above] {$h_i$};
			\draw [-] (-0.9,0.0) --  (-0.9,2.2);
			\draw [|<->|, dashed] (-2.5,0.2) --  (-4.2,0.2) node[left] {$r_i=\sum_{\substack{j=1 \\ j \ne i}}^{n} |a_{ij}| $};
			\draw [-] (-0.9,0.0) --  (-0.9,2.2);
			\draw[scale=1.0,domain=-3.141:3.141,smooth,variable=\t]
			plot ({-2.5+1.6*sin(\t r)},{1.6*cos(\t r)});
			\end{tikzpicture}       
		\end{minipage}\hfill
		\begin{minipage}{0.50\textwidth}
			\centering
			\begin{tikzpicture}
			\draw[] (-7, 3.5) node[above] {(B)};
			
			\draw[->] (-4.2,0) -- (1.4,0) node[right] {$Re$};
			\draw[->] (0,-2.2) -- (0,3.50) node[above] {$Im$};
			
			\draw[->] (0.0,0.0) -- (-2.3,2.5) node[above] {$|z - a_{ii}|$};
			\draw[->] (-1.5,0.0) -- (-2.3,2.5);
			\draw [|<->|, dashed] (0,-0.2) --  (-1.5,-0.2) node[below] {$a_{ii} < 0$};
			\draw [|<->|, dashed] (0,2.2) --  (1.1,2.2) node[above] {$h_i$};
			\draw [|<->|, dashed] (-1.5,0.2) --  (-4.1,0.2) node[left] {$r_i=\sum_{\substack{j=1 \\ j \ne i}}^{n} |a_{ij}|$};
			\draw [-] (1.1,0.0) --  (1.1,2.2);
			\draw[scale=1.0,domain=-3.141:3.141,smooth,variable=\t]
			plot ({-1.5+2.6*sin(\t r)},{2.6*cos(\t r)});
			\end{tikzpicture}        
		\end{minipage}
		\hspace*{\fill}
	\end{minipage}
	\begin{minipage}{1\textwidth}
		\begin{minipage}{0.50\textwidth}
			\centering
			\begin{tikzpicture}[domain=0:4]
			\draw[] (-1.5, 3.8) node[above] {(C)};
			
			\draw[->] (-0.8,0) -- (4.3,0) node[below] {$Re$};
			\draw[->] (0,-2.2) -- (0,3.50) node[above] {$Im$};
			
			\draw[->] (0.0,0.0) -- (2.9,1.5) node[above] {$|z - a_{ii}|$};
			\draw[->] (2.5,0.0) -- (2.9,1.5);
			\draw [|<->|, dashed] (0,-0.2) --  (2.5,-0.2) node[below] {$a_{ii} > 0 $};
			\draw [|<->|, dashed] (0,2.2) --  (0.9,2.2) node[above] {$h_i$};
			\draw [-] (0.9,0.0) --  (0.9,2.2);
			\draw [|<->|, dashed] (2.5,0.3) --  (4.1,0.3) node[right] {$r_i=\sum_{\substack{j=1 \\ j \ne i}}^{n} |a_{ij}| $};
			\draw [-] (0.9,0.0) --  (0.9,2.2);
			\draw[scale=1.0,domain=-3.141:3.141,smooth,variable=\t]
			plot ({2.5+1.6*sin(\t r)},{1.6*cos(\t r)});
			\end{tikzpicture}       
		\end{minipage}\hfill
		\begin{minipage}{0.50\textwidth}
			\centering
			\begin{tikzpicture}[domain=0:4]
			\draw[] (-2, 3.8) node[above] {(D)};
			
			\draw[->] (-1.2,0) -- (4.2,0) node[below] {$Re$};
			\draw[->] (0,-2.2) -- (0,3.50) node[above] {$Im$};
			
			\draw[->] (0.0,0.0) -- (2.1,2.4) node[above] {$|z - a_{ii}|$};
			\draw[->] (1.5,0.0) -- (2.1,2.4);
			\draw [|<->|, dashed] (0,-0.2) --  (1.5,-0.2) node[below] {$a_{ii} > 0$};
			\draw [|<->|, dashed] (0,2.2) --  (-1.0,2.2) node[above] {$h_i$};
			\draw [|<->|, dashed] (1.5,0.2) --  (4.0,0.2) node[right] {$r_i=\sum_{\substack{j=1 \\ j \ne i}}^{n} |a_{ij}|$};
			\draw [-] (-1.0,0.0) --  (-1.0,2.2);
			\draw[scale=1.0,domain=-3.141:3.141,smooth,variable=\t]
			plot ({1.5+2.5*sin(\t r)},{2.5*cos(\t r)});
			\end{tikzpicture}        
		\end{minipage}
		\hspace*{\fill}
	\end{minipage}
	\caption{\textbf{Types pf Gershgorin's circles for eigenvalues:} (A) Type 1. (B) Type 2. (C) Type 3. (D) Type 4.}
	\label{fig:1}
\end{figure}
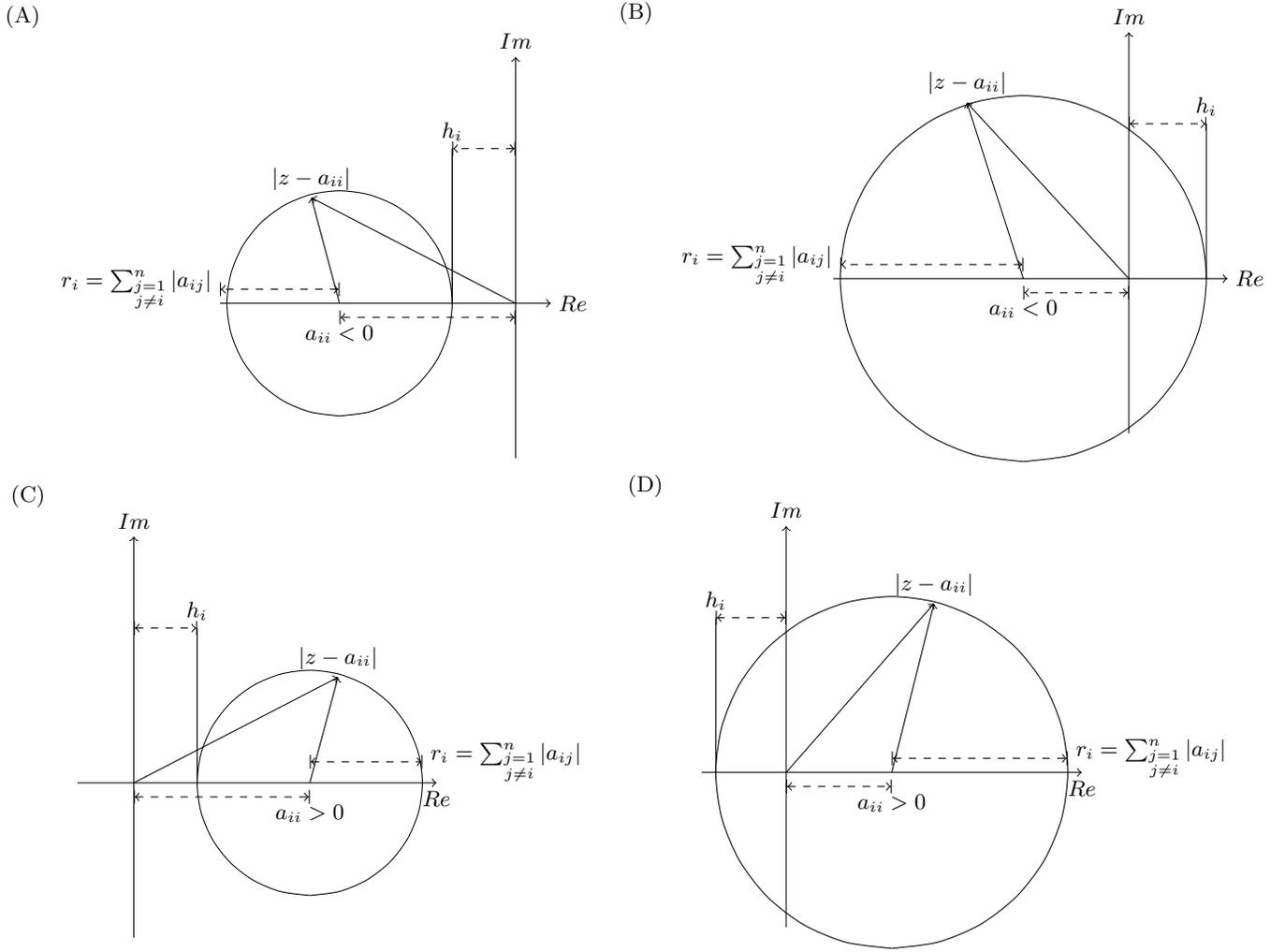

\subsection{Overlapping circles}
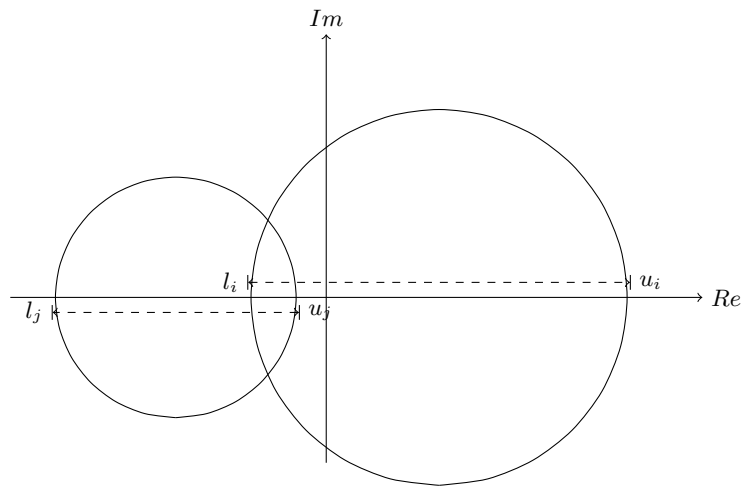
\begin{figure}
	\centering
	\begin{tikzpicture}[domain=0:4]
	\draw[->] (-4.2,0) -- (5.0,0) node[right] {$Re$};
	\draw[->] (0,-2.2) -- (0,3.50) node[above] {$Im$};
	\draw[scale=1.0,domain=-3.141:3.141,smooth,variable=\t]
	plot ({1.5+2.5*sin(\t r)},{2.5*cos(\t r)});
	\draw[scale=1.0,domain=-3.141:3.141,smooth,variable=\t]
	plot ({-2.0+1.6*sin(\t r)},{1.6*cos(\t r)});
	
	\draw [|<->|, dashed] (-1.05,0.2) node[left] {$l_i$} --  (4.05,0.2) node[right] {$u_i$};
	
	\draw [|<->|, dashed] (-0.35,-0.2) node[right] {$u_j$} --  (-3.65,-0.2) node[left] {$l_j$};
	
	\end{tikzpicture}
	\caption{\textbf{Two overlapping circles.}}
	\label{fig:inter}
\end{figure}

We must emphasize that when two or more Gershgorin's circles overlap, the eigenvalues lie in the union of the circles. For instance, in Fig. \ref{fig:inter}, the union of two circles shows that the real part of both eigenvalues corresponding to rows $i$ and $j$ must be in the union of their diameters
\begin{equation}
Re(\lambda_i), Re(\lambda_j) \in [l_j, u_j] \bigcup [l_i, u_i] = [l_j, u_i].
\end{equation}

The lower and upper bounds of each circle must be

\begin{equation}
l_i \equiv Re(a_{ii}) - \sum_{\substack{j=1 \\ j \ne i}}^{n} |a_{ij}| = Re(a_{ii}) - r_i,
\end{equation}
and
\begin{equation}
u_i \equiv Re(a_{ii}) + \sum_{\substack{j=1 \\ j \ne i}}^{n} |a_{ij}| = Re(a_{ii}) + r_i,
\end{equation}
respectively, and for each row (or column) of the matrix $A$, there exists a corresponding interval $[l_i, u_i]$ 

\begin{equation}
A= \begin{pmatrix}
a_{11} & a_{12} & \dots & a_{1n}\\
a_{21} & a_{22} & \dots & a_{2n}\\
\vdots & \vdots & \ddots & \vdots \\
a_{n1} & a_{n2} & \dots & a_{nn}\\
\end{pmatrix}
\implies
\begin{matrix}
[l_1, u_1] \\
[l_2, u_2] \\
\vdots \\
[l_n, u_n] \\
\end{matrix},
\end{equation}
where after taking their unions, one or more disjoint intervals results in 

\begin{equation}
\mathcal{A} = \bigcup_{i=1}^{n} [l_i, u_i] \equiv \bigcup_{i=1}^{m}  [L_i, U_i] , \qquad 1 \le m  \le n,
\end{equation}
such that 

\begin{equation}
[L_i, U_i] \cup [L_j, U_j] = \emptyset, \quad i \ne j.
\end{equation}
After taking the unions, we can see that $L_i$s and $U_i$s are not necessarily equal to $l_i$s and $u_i$s. Still by defining
\begin{equation}
U_{max} \equiv \max \{ U_i: i = 1, \dots, m\} = \max \{ u_i: i = 1, \dots, m\},
\end{equation}
the $[L_{max}, U_{max}]$ corresponds to the right-most disjoint interval.
Recall that our main goal in studying the linear stability of a system is investigating its Jacobian eigenvalues, and if the real part of the largest eigenvalue is positive, one concludes that the system is unstable. Similarly, the negative real part of the largest eigenvalue implies the stability of the system. 

As a result, the right-most disjoint interval in $\mathcal{A}$ is enough to conclusively state the sign of the largest eigenvalue, and $[L_{max}, U_{max}]$ introduces the following possibilities regarding the stability condition:
\begin{enumerate}
	\item For $U_{max} \le 0$, the real part of the largest eigenvalue is negative. Consequently, the system is stable.
	\item For $L_{max} > 0$, the real part of the largest eigenvalue is positive. Consequently, the system is unstable.
	\item  For $L_{max} \le 0 < U_{max}$ the situation is inconclusive.
\end{enumerate}
All three cases are depicted in Fig. \ref{fig:3}A, \ref{fig:3}B and \ref{fig:3}C.

\begin{figure}
	\centering
	\begin{minipage}{0.3\textwidth}
		\centering
		\begin{tikzpicture}[domain=0:4]   
		\draw[] (-3.2,1.7) node[above] {(A)};
		\draw[->] (-3.0,0) -- (0.8,0) node[right] {$Re$};
		\draw[->] (0,-1.5) -- (0,1.50) node[above] {$Im$};        
		\draw [|-|, line width=0.2mm] (-2.5,-0.0) node[below] {$L_{max}$} --  (-0.5,-0.0) node[below] {$U_{max}$}; 
		\draw [-, line width=1.0mm] 
		(-2.5,-0.01) --  (-0.5,-0.01); 
		\end{tikzpicture}        
	\end{minipage}\hfill
	\begin{minipage}{0.3\textwidth}
		\centering
		\begin{tikzpicture}[domain=0:4] 
		\draw[] (-1.0,1.7) node[above] {(B)};
		\draw[->] (-0.8,0) -- (3.0,0) node[right] {$Re$};
		\draw[->] (0,-1.5) -- (0,1.50) node[above] {$Im$};        
		\draw [|-|, line width=0.2mm] (0.5,-0.0) node[below] {$L_{max}$} --  (2.5,-0.0) node[below] {$U_{max}$}; 
		\draw [-, line width=1.0mm] 
		(0.5,-0.01) --  (2.5,-0.01); 
		\end{tikzpicture}
	\end{minipage}\hfill
	\begin{minipage}{0.3\textwidth}
		\centering
		\begin{tikzpicture}[domain=0:4] 
		\draw[] (-2.0,1.7) node[above] {(C)};
		\draw[->] (-1.8,0) -- (1.5,0) node[right] {$Re$};
		\draw[->] (0,-1.5) -- (0,1.50) node[above] {$Im$};        
		\draw [|-|, line width=0.2mm] (-1.5,-0.0) node[below] {$L_{max}$} --  (.5,-0.0) node[below] {$U_{max}$}; 
		\draw [-, line width=1.0mm] 
		(-1.5,-0.01) --  (.5,-0.01); 
		\end{tikzpicture}
	\end{minipage}\hspace*{\fill}
	\caption{\textbf{Right-most union's stability conditions}. (A) $U_{max} \le 0$. (B) $L_{max} > 0$. (C) $L_{max} < 0$ and $U_{max} \ge 0$.}
	\label{fig:3}
\end{figure}
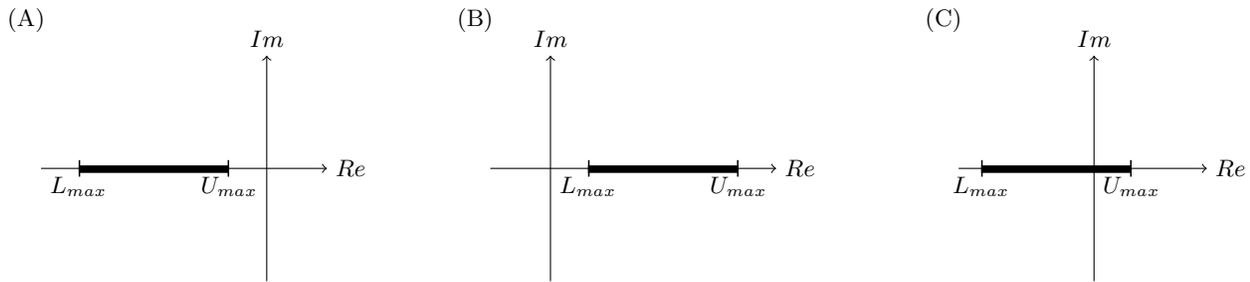

\subsection{Linear Stability Analysis}
At this stage, we can use the obtained results in two different ways. The first possibility is when the Jacobian is written in terms of the model's parameters, $\bm{\theta}$, and we may be able to derive the right-most disjoint set parametrically. Accordingly, the inequalities mentioned in the previous section define the regions in parameter space where the method can conclusively determine the stability/instability of the system. Nevertheless, the region corresponding to the inconclusive range requires different classification techniques. We used this technique for Brusselator \cite{Prigogine1985} (see the main paper) and Lengyel-Epstein \cite{Lengyela1990,Lengyelb1990} (see subsection \ref{sec:lengyel-epstein}) model. Note that even if the method is not always conclusive for all the regions in parameter space, it can always find a theoretical lower bound for the volume of the parameter space that the system is stable/unstable.

Furthermore, the second possibility is when studying a system's linear stability numerically. We propose the Algorithm (\ref{alg:1}) that classifies a given Jacobian matrix into ``\textit{stable}'', ``\textit{unstable}'' and ``\textit{inconclusive}'' groups. 

\newpage
\section{Reaction-diffusion models}
To study a pattern-forming system like the reaction-diffusion model, for instance, in 
\begin{equation}
\label{eq:reaction-diffusion}
\frac{d {X}_i}{dt} = D_i \nabla^2 X_i + f_i(X_1, \dots, X_n), \quad i = 1, \dots, n,
\end{equation}
after linearising equations at the stationary solution $\bm{X}^* = (X^*_1, \dots, X^*_n)$, the Jacobian writes as 
\begin{equation}
\left. \bm{J} \right|_{\bm{X}^*} = \begin{pmatrix}
\partial_1 f_1 & \partial_2 f_1 & \dots & \partial_n f_1 \\
\partial_1 f_2 & \partial_2 f_2 & \dots & \partial_n f_2 \\
\vdots & \vdots & \ddots  & \vdots \\
\partial_1 f_n & \partial_2 f_n & \dots & \partial_n f_n \\
\end{pmatrix}.
\end{equation}
where $\partial_i = \partial / \partial u_i$. Thus, to study the linear stability of the matrix $\bm{J}$ by using the proposed method, we can write the lower and upper bounds corresponding to row (column) $i$ as
\begin{equation}
l_i = \partial_i f_i - r_i,
\end{equation}
and
\begin{equation}
u_i = \partial_i f_i + r_i,
\end{equation}
for $r_i = \sum_{\substack{j=1 \\ j \ne i}}^{n} |\partial_j f_i|$.

Then, the stability/instability criteria of the Jacobian determine the Turing pattern conditions. For unstable cases, including diffusion will not produce a Turing pattern; therefore, there is no need to complete linear stability analysis. However, it is interesting to see the effect of introducing diffusion for inconclusive and stable cases. 

\begin{figure}
	\centering
	\begin{tikzpicture}
	
	\draw[->] (-8.2,0) -- (1.4,0) node[right] {$Re$};
	\draw[->] (0,-2.2) -- (0,3.50) node[above] {$Im$};
	
	\draw[->] (0.0,0.0) -- (-2.3,2.5) node[above] {$|z - \partial_i f_i|$};
	\draw[->] (-1.5,0.0) -- (-2.3,2.5);
	\draw [|<->|, dashed] (0,-0.2) --  (-1.5,-0.2) node[left] {$\partial_i f_i$};          
	\draw[scale=1.0,domain=-3.141:3.141,smooth,variable=\t]
	plot ({-1.5+2.6*sin(\t r)},{2.6*cos(\t r)});
	
	\draw[scale=1.0,domain=-3.141:3.141,smooth,variable=\t]
	plot ({-5.0+2.6*sin(\t r)},{2.6*cos(\t r)});
	\draw [|<->|, dashed] (0,0.2) --  (-5.0,0.2) node[above] {$\partial_i f_i -k^2 D_i$}; 
	\draw [->, line width=.5 mm] (-1.5,-2.6) --  (-5.0,-2.6) node[below] { $-k^2 D_i$}; 
	
	\draw [->, line width=.5 mm] (-1.5,-2.6) --  (-5.0,-2.6) node[below] { $-k^2 D_i$}; 
	\draw [dashed] (-5.0,0) --  (-5.0,-2.6);
	\draw [dashed] (-1.5,0) --  (-1.5,-2.6);
	\end{tikzpicture}
	\caption{{\bf Effect of diffusion on circles}. The centre of the circle shifts to the left by an amount $-k^2 D_i$.}
	\label{fig:shift}
\end{figure}
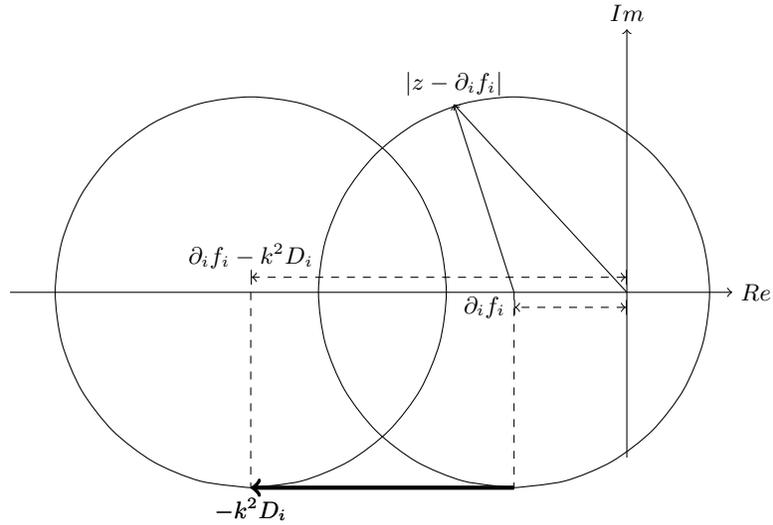

For a given wave number $k$, the inclusion of diffusion shifts all the diagonal terms by $-k^2 D_i$ accordingly 

\begin{equation}
\boldsymbol{J} (k) = \left. \boldsymbol{J}_0 \right|_{\boldsymbol{X}^*} - k^2 \bm{D}.
\end{equation}
Effectively, since diagonal terms correspond to the location of Gershorin's circle and only diagonal terms are affected by including the diffusion, it is geometrically equivalent to saying all the centre of Gershgorin's circle shifts to the left. Simultaneously, since the off-diagonal terms have not changed, the circle's radius remains unchanged. This effect is shown in Fig. \ref{fig:shift}.

Indeed, for any circle with a segment on the positive side of the real axis, there exists a maximum shift to the left, say $k^{*2}_i D_i$, by a wave number denoted by $k^{*}_i$ that transfers the circle entirely to the negative side of the real axis. Also, after the shift, the real part of the eigenvalue corresponding to that circle must be negative for all the higher wave numbers $k_i > k^{*}_i$. Hence,  depending on the sign of $\partial_i f_i$ the $k^{*}_i$ derives as
\[
\begin{cases}
r_i - |\partial_i f_i| - k^{*2}_i D_i = 0, & \partial_i f_i \le 0\\
\\
r_i + |\partial_i f_i| - k^{*2}_i D_i = 0, & \partial_i f_i > 0
\end{cases} \implies
\]
\begin{equation}
k^{*}_{i} = \sqrt{\frac{r_i + \partial_i f_i}{D_i}},
\end{equation}
for $r_i = \sum_{\substack{j=1 \\ j \ne i}}^{n} |\partial_j f_i|$.

Subsequently, the linear stability of different wave numbers restricts to the range $k_i \in [0, k^{*}_{i}]$, such that for $k=0$, or the case with no diffusion, $\left. \boldsymbol{J}_0 \right|_{\boldsymbol{X}^*}$ determines the stability of the initial condition and $\boldsymbol{J}(k)$ for $k_i \in (0, k^{*}_{i}]$ determines that the evolution of the dominant wave number in a perturbed system:

\begin{itemize}
	\item For $\left. \boldsymbol{J}_0 \right|_{\boldsymbol{X}^*}$, when $U_{max} \le 0$, all the circles are on the negative side of the real axis and including the diffusion for any given wave number shifts the circles further to the left by an amount $-k_i^2 D_i$. Hence, all real parts of eigenvalues are negative, and diffusion cannot excite any wave number. Consequently, the system is incapable of producing a Turing pattern (``\textit{super-stable}'').
	
	\item On the contrary, when $U_{max} > 0$ for $\left. \boldsymbol{J}_0 \right|_{\boldsymbol{X}^*}$, the initial stationary state is unstable and incapable of producing a Turing pattern (``\textit{unstable}'').
	
	\item Finally, when $[L_{max}, U_{max}]$ is on inconclusive range, $\left. \boldsymbol{J}_0 \right|_{\boldsymbol{X}^*}$ must be studied by finding its eigenvalues. And if one finds that it is stable, the maximum of the dispersion relation $\lambda(k)$ finds the dominant wave number for pattern formation. Meanwhile, by finding $k^{*}_{max} \equiv \max \{k^{*}_{1}, \dots, k^{*}_{n} \}$ for all rows (or columns), the study of the dispersion relation can be restricted to $[0, k^{*}_{max}]$. 
\end{itemize}

Similar to the stability analysis of a single matrix, Algorithm (\ref{alg:1}) can speed up the process to check the possibility of a Turing pattern for a given parameter set. However, the algorithm can further speed up the search for Turing pattern-formation systems. 

To elaborate, let us recite the linear stability of the pattern-formations models: Without diffusion, the $\left. \boldsymbol{J}_0 \right|_{\boldsymbol{X}^*}$ must be stable, and after the inclusion of diffusion, $\bm{J}(k)$ must be unstable for some $k$s. In contrast, when algorithm (\ref{alg:1}) classifies $\left. \boldsymbol{J}_0 \right|_{\boldsymbol{X}^*}$ as stable, it is impossible to have a pattern as a stationary solution (This is why we called that class ``\textit{super-stable}''). The pattern-forming parameters are among those that are classified as inconclusive and need further study by their eigenvalues.

\begin{remark}
	The dispersion relation for all diffusing species always satisfies
	\begin{equation}
	k > k^{*}_{max}: Re(\lambda(k)) < 0.
	\end{equation}
	In other words, asymptotically, as long as all species are diffusers, we have
	\begin{equation}
	\lim_{k \rightarrow \infty} Re(\lambda(k)) \rightarrow -\infty.
	\end{equation}
\end{remark}

\newpage
\section{Tightening the bounds}
Given an invertible matrix $\bm{D}$, observe that $\bm{B} = \bm{D} \bm{A} \bm{D}^{-1}$ introduces an equivalence relation between square matrices $\bm{A}$ and $\bm{B}$ such that the matrix $\bm{B}$ have the same eigenvalues as $\bm{A}$. To see that, let us assume $\lambda$ and $\bm{x}$ are $\bm{A}$'s eigenvalue and its corresponding eigenvector
\begin{equation}
\bm{A} \bm{x} = \lambda \bm{x}.
\end{equation}
Then,
\begin{eqnarray}
\bm{B} (\bm{D} \bm{x}) &=& (\bm{B} \bm{D}) \bm{x} \nonumber \\
&=& (\bm{D} \bm{A}) \bm{x} \nonumber \\
&=& (\bm{D} ) \lambda \bm{x} \nonumber \\
&=& \lambda (\bm{D} \bm{x}).
\end{eqnarray}
which shows $\lambda$ is the eigenvalue of $\bm{B}$ with the corresponding eigenvector $\bm{D} \bm{x}$.
Observing that, we define an invertible $n \times n$ diagonal matrix $\bm{D}$ as
\begin{equation}
\bm{D} = 
\begin{pmatrix}
1  & \dots & 0 & 0 & 0  &\dots & 0 \\        
\vdots & \ddots & \quad & \vdots & \quad  & \quad & \vdots \\
0   & \dots& 1 & 0 & 0  &\dots & 0 \\
0  & \dots & 0 & \frac{1}{d_i} & 0  &\dots & 0 \\
0   & \dots & 0 & 0 & 1  &\dots & 0 \\
\vdots & \quad & \quad & \vdots & \quad & \ddots  & \vdots\\
0 & \dots & 0 & 0 & 0  &\dots &  1\\
\end{pmatrix},
\end{equation}
the transformed matrix $\bm{D} A \bm{D}^{-1}$ has the form
\begin{eqnarray}
&
\begin{pmatrix}
1  & \dots & 0 &\dots & 0 \\        
\vdots & \ddots & \vdots & \quad & \vdots \\       
0  & \dots & \frac{1}{d_i} &\dots & 0 \\
\vdots & \quad & \vdots & \ddots & \vdots\\
0 & \dots & 0 &\dots &  1\\
\end{pmatrix}
\begin{pmatrix}
a_{11}  & \dots & a_{1i} &\dots & a_{1n} \\        
\vdots & \ddots & \vdots & \quad & \vdots \\       
a_{i1}  & \dots & a_{ii} &\dots & a_{in} \\
\vdots & \quad & \vdots & \ddots & \vdots \\
a_{n1} & \dots & a_{ni} &\dots &  a_{nn}\\
\end{pmatrix}
\begin{pmatrix}
1  & \dots & 0 &\dots & 0 \\        
\vdots & \ddots & \vdots & \quad & \vdots \\       
0  & \dots & d_i &\dots & 0 \\
\vdots & \quad & \vdots & \ddots & \vdots \\
0 & \dots & 0 &\dots &  1\\
\end{pmatrix} &  \nonumber \\
=& \begin{pmatrix}
a_{11}  & \dots & a_{1i} d_i &\dots & a_{1n} \\        
\vdots & \ddots & \vdots & \quad & \vdots \\       
\frac{a_{i1}}{d_i}  & \dots & a_{ii} &\dots & \frac{a_{in}}{d_i} \\
\vdots & \quad & \vdots & \ddots & \vdots \\
a_{n1} & \dots & a_{ni} d_i &\dots &  a_{nn}\\
\end{pmatrix}, &
\end{eqnarray}
the transformation's effect is similar to dividing all the elements of the row $i$ by $d_i$ and multiplying the elements of the column $i$ by $d_i$. Consequently, the diagonal term $a_{ii}$ remains the same.

Consider the rows of the resulting matrix. The interval of all rows except the $i$-th will change from
\begin{eqnarray}
[l_j, u_j] &=& 
\left[a_{jj} -  r_j, \quad a_{jj} +  r_j \right]   \nonumber \\
&=& 
\left[a_{jj} -  \sum_{\substack{k=1 \\ k \ne j}}^{n} |a_{jk}|, \quad a_{jj} +  \sum_{\substack{k=1 \\ k \ne j}}^{n} |a_{jk}|\right],
\end{eqnarray}
to
\begin{eqnarray}
\label{eq:d_j_bounds2}
[l_j, u_j] &=& 
\left[a_{jj} - |a_{ji}| d_i -  \sum_{\substack{k=1 \\ k \ne i, j}}^{n} |a_{jk}| , \quad a_{jj} + |a_{ji}| d_i +  \sum_{\substack{k=1 \\ k \ne i,j}}^{n} |a_{jk}| \right] \nonumber \\
&=& \left[a_{jj} - |a_{ji}| (d_i-1) -  r_j , \quad a_{jj} + |a_{ji}| (d_i-1) + r_j \right] 
,    
\end{eqnarray}
and the row $i$ to
\begin{eqnarray}
\label{eq:d_i_bounds2}
[l_i, u_i] &=& 
\left[a_{ii} -  \sum_{\substack{k=1 \\ k \ne i}}^{n} \frac{|a_{ik}|}{d_i} , \quad a_{ii}  +  \sum_{\substack{k=1 \\ k \ne i}}^{n} \frac{|a_{ik}|}{d_i} \right]    \nonumber \\
&=& 
\left[a_{ii} -   \frac{r_i}{d_i} , \quad a_{ii}  +  \frac{r_i}{d_i} \right]
\end{eqnarray}

In other words, the radii of all circles correspond to rows other than $i$ expand by the amount $|a_{ji}| (d_i-1)$ and the $i$-th radius shrinks by the factor of $1/d_i$, while the centre of all circles stays on the same point.
And since the eigenvalues of the transformed matrix are the same as the original one, one can hope the shrunk circle becomes isolated from the rest. At the same time, the other radius expansions were smaller than that single shrink. In practice, one can find $d_i$ that can isolate the circle with the largest centre from the rest.
To tighten the bounds, we have two distinct cases that study separately:  
\begin{itemize}
	\item Let us assume $a_{ii} > 0$ is the largest diagonal term. To isolate the circle corresponding to row $i$, its leftmost point, or $l_i = a_{ii} -  \frac{r_i}{d_i}$, must be larger than every other circle's rightmost point, or $u_j =  a_{jj} + |a_{ji}| (d_i-1) +  r_j$. For a given row $j$, this condition writes as
	\[
	a_{jj} + |a_{ji}| (d_i-1) +  r_j <
	a_{ii} -  \frac{r_i}{d_i} \implies
	\]
	\begin{equation}
	\label{eq:quad_d}
	|a_{ji}| d^2_i + (a_{jj} - a_{ii} +  r_j - |a_{ji}|) d_i + r_i < 0.    
	\end{equation}
	The last result is a quadratic form in $d_i$ with all its three coefficients calculated for the Algorithm (\ref{alg:1}) for the rows $i$ and $j$. To satisfy the inequality, the discriminant of the quadratic form must be positive, or
	\begin{equation}
	\label{eq:shrink-cond-1}
	(a_{jj} - a_{ii} +  r_j - |a_{ji}|)^2 > 4 |a_{ji}| r_i.  
	\end{equation}
	At the same time, since $r_i >0$, to have one or more positive solutions for the quadratic equation, say $d_i >0$, we must have
	\begin{equation}
	\label{eq:shrink-cond-2}
	a_{jj} - a_{ii} +  r_j - |a_{ji}| < 0.
	\end{equation}
	And finally, $l_i = a_{ii} - r_i/d_i$ must be positive to have an unstable Jacobian, which constraint $d_i$ as
	\begin{equation}
	\label{eq:shrink-cond-3}
	\frac{r_i}{a_{ii}} < d_i.
	\end{equation}
	To check this condition, one needs to find the roots of  the quadratic Eq. (\ref{eq:quad_d}) and check that its largest solution is greater than $r_i/a_{ii}$. If all the rows $j$ and the largest one, $i$, satisfy the inequalities 
	\begin{eqnarray}
	\begin{cases}    
	(a_{jj} - a_{ii} +  r_j - |a_{ji}|)^2 > 4 |a_{ji}| r_i\\
	a_{jj} - a_{ii} +  r_j - |a_{ji}| < 0  \\ 
	\frac{r_i}{a_{ii}} < d_i
	\end{cases}
	\begin{split}
	j &= 1, \dots, n\\    
	j &\ne i
	\end{split},
	\end{eqnarray}
	there exists a $d_i$ that isolates the rightmost eigenvalue, and the Jacobian is conclusively unstable.
	
	\item When the largest diagonal term is negative, or  $a_{ii} < 0$, it implies all the other diagonal terms are negative two. In this case, we search for a possible shrinkage value $d_i$, such that while the circle of the row $i$ shrinks with its upper bound on the negative side of the real axis, the growth of all the other circles keeps them at the negative side too. This argument is equivalent to a set of conditions for the row $i$ as
	\begin{equation}
	\frac{r_i}{|a_{ii}|} < d_i,
	\end{equation}
	and for all the other rows as
	\begin{equation}
	d_i < \frac{|a_{jj}| - r_j}{|a_{ji}|} + 1.
	\end{equation}
	Combining these conditions derives a bound as
	\begin{equation}
	\frac{r_i}{|a_{ii}|} < d_i < \min_{j \ne i} (\frac{|a_{jj}| - r_j}{|a_{ji}|} + 1).
	\end{equation}
	So, if a non-empty interval can be found that satisfies the above inequalities, Jacobian is conclusively super-unstable. 
\end{itemize}

Using these results, we propose the Algorithm (\ref{alg:2}) that can search among the inconclusive cases from the classification of the Algorithm (\ref{alg:1}) for classifying more cases into super-stable, unstable and inconclusive.

\newpage
\section{Algorithms}
\subsection{Algorithm \ref{alg:1}}
\begin{algorithm}
	\caption{Search algorithm for finding the union of rows (or columns) to classify  the stability of a matrix.}
	\label{alg:1}
	\begin{algorithmic}
		\State $U_{max} \gets \max \{ u_1, \dots, u_n \}$
		
		\State $i \gets$ index$(U_{max})$
		\Comment{The index of the maximum upper bound}
		\State $L_i \gets [l_{i}, \quad u_{i}]$
		\Comment{The corresponding lower bound}
		
		\IF{$U_{max} < 0$} \RETURN{ Super-stable} 
		\ENDIF
		
		\IF{$L_i < 0$}
		\RETURN{ Inconclusive}
		\ENDIF
		
		\FOR {$j \in \{1, \dots, n\}/\{i\}$}
		\State $L_j, U_j \gets [l_i, u_i]$
		\IF{$L_i < U_j$}
		\IF{$L_j < L_i$}
		\State $L_j \gets L_i$
		\ENDIF
		\IF{$L_i < 0$}
		\RETURN{ Inconclusive}
		\ENDIF         
		\ENDIF   
		\ENDFOR
		
		\RETURN{ Unstable}
		\\\hrulefill
	\end{algorithmic}
\end{algorithm}
\subsection{Algorithm \ref{alg:2}}
\begin{algorithm}
	\caption{Search algorithm for finding the tightening bounds to classify the stability of a matrix.}\label{alg:2}
	\begin{algorithmic}
		
		\State $a_{max} \gets \max \{ a_{11}, \dots, a_{nn} \}$ \Comment{The largest diagonal term}
		
		\State $i \gets$ index$(a_{max})$ \Comment{Index of the largest diagonal term}

		\IF{$a_{max} > 0$}

		\FOR {$j \in \{1, \dots, n\}/\{i\}$}
		
		$d_{max} \gets \max\{Roots(|a_{ji}| d^2_i + (a_{jj} - a_{ii} +  r_j - |a_{ji}|) d_i + r_i)\}$ \Comment{The largest roots of the equation}
		\IF {$d_{max} \le r_i/a_{max}$}
		\RETURN{ Inconclusive}
		\ENDIF
		\IF {$(a_{jj} - a_{ii} +  r_j - |a_{ji}|) \ge 0$}
		\RETURN{ Inconclusive}
		\ENDIF
		\IF {$(a_{jj} - a_{ii} +  r_j - |a_{ji}|)^2 \le 4 |a_{ji} r_i|$}
		\RETURN{ Inconclusive}
		\ENDIF   
		\ENDFOR
		
		\RETURN{ Unstable}
		\ELSE
		
		\IF{$r_i/|a_{ii}| \ge \min_{j \ne i} \{\frac{|a_{jj}| - r_j}{|a_{ji}|} + 1 \}, \quad j \in \{1, \dots, n\}/\{i\} $} \RETURN{ Inconclusive}
		\ENDIF
		
		\RETURN{ Super-stable}
		
		\ENDIF
		\\\hrulefill
	\end{algorithmic}
\end{algorithm}

\newpage
\section{Results}
\subsection{Numerical comparison}
\label{sec:numerical-comp}
To test the ratio of rejection and, consequently, the speed up due to the algorithm  (\ref{alg:1}), we use a biologically inspired model capable of producing a Turing pattern \cite{scholes_2019}. The model reaction-diffusion PDE with nine free parameters is written as 
\begin{eqnarray}
\frac{\partial u}{\partial t} &=& b_u  +     \frac{V_u}{\left[1+\left(\frac{K_{uu}}{u }\right)^{4}\right] \left[1+\left(\frac{v}{K_{vu} }\right)^{4}\right]} 
\nonumber \\
&&-  \mu_u u + D_u \nabla^2 u, \nonumber\\
\frac{\partial v}{\partial t} &=& b_v  +     \frac{V_v}{1+\left(\frac{K_{uv}}{u }\right)^{4}}  -  \mu_v v + D_v \nabla^2 v.
\label{eq:Hill-function-pde}
\end{eqnarray}
Note that the nonlinear terms are Hill functions that regulate activation and inhibition in a cell's gene expression. The Jacobian of the linearised form of the above equations is a two-by-two matrix, and in practice, the computational cost of calculating its eigenvalues is not much different than our algorithm. However, we selected this model for the numerical experiment since it is easy to check the stability/instability by comparing the determinant and trace of the Jacobian and ensuring the algorithm's correctness.

In this simulation, we used diffusion constants $D_u = 0.01$ and $D_v = 1$ and selected all one billion parameter combinations from the nine-dimensional mesh grid  $\{0.01, 0.05, 0.1, 0.5, 1, 5, 10, 50, 100, 500\}^9$ and applied the algorithms (\ref{alg:1}) (\ref{alg:2}) and to classify them into ``\textit{unstable}'', ``\textit{super-stable}'', ``\textit{Inconclusive}'' and ``\textit{No fixed point}''. Note that the case ``No fixed point'' refers to the parameter combinations that the root finding algorithm (``hybrd'' of ``MINPACK'' library) could not find a stationary solution. We first used the Algorithm (\ref{alg:1}) for row-wise comparison, and after that, by using the inconclusive results from the first run, we used the algorithm again for column-wise calculation. And finally, classified the remaining inconclusive cases by using the Algorithm (\ref{alg:2}). These results are presented in table (\ref{table:1}).

\begin{table}[h!]
	\begin{center}
		\begin{tabular}{ |c|c|c|c|c|c| } 
			\hline
			& \textbf{Total} & \textbf{Super-stable} & \textbf{Inconclusive} & \textbf{Unstable} & \textbf{No fixed point}\\ 
			\hline
			\multirow{2}{6em}{Row-wise} & $10^9$  & $850,677,030$ & $140,394,311$ & $4,870,615$ & $4,058,044$ \rule{0pt}{2.6ex} \\ 
			
			& 100\%  & 85.07\% & 14.04\% & 0.49\% & 0.41\% \rule{0pt}{2.6ex} \\ 
			\hline
			\multirow{2}{6em}{Column-wise}  & $140,394,311$  & $68,454,498$ & $71,913,845$ & $25,968$ & - \rule{0pt}{2.6ex} \\ 
			
			& 100\%  & 48.76\% & 51.22\% & 0.02\% & - \rule{0pt}{2.6ex} \\ 
			\hline
			\multirow{2}{6em}{Tighten bounds}  & $71,913,845$  & $64,615,611$ & $6,990,298$ & $307,936$ & - \rule{0pt}{2.6ex} \\ 
			
			& 100\%  & 89.85\% & 9.72\% & 0.43\% & - \rule{0pt}{2.6ex} \\ 
			\specialrule{.2em}{.1em}{.1em} 
			\multirow{2}{6em}{Combined} & $10^9$  & $983,747,139$ & $6,990,298$ & $5,204,519$ & $4,058,044$ \rule{0pt}{2.6ex} \\ 
			
			& 100\%  & 98.37\% & 0.70\% & 0.52\% & 0.41\% \rule{0pt}{2.6ex} \\ 
			\hline
		\end{tabular}
		\caption{Algorithm (\ref{alg:1}) empirical statistics for the parameter space search.}
		\label{table:1}
	\end{center}
\end{table}

The table (\ref{table:1}) shows that more than $99.3\%$ of the parameter combinations were rejected. Even in this case, where calculating the eigenvalues is computationally cheap, there is a substantial speed up in comparison to the usual Turing space search: the rejected super-stable cases do not need a dispersion relation study since we know they will stay stable even after diffusion inclusion.

\subsection{Lengyel-Epstein model}
\label{sec:lengyel-epstein}
The Lengyel-Epstein model \cite{Lengyela1990, Lengyelb1990} is written for two species, $u$ and $v$ like
\begin{equation}
\frac{d u}{d t} = \frac{1}{\sigma}\left( \nabla^2 u + a - u - 4 \frac{uv}{1 + u^2}\right),
\end{equation}
and
\begin{equation}
\frac{d v}{d t} = d  \nabla^2 v + b \left(u - \frac{uv}{1 + u^2}\right),
\end{equation}
for parameters $\sigma, d, a, b > 0$. The Jacobian matrix for the stationary state 
\begin{equation}
(u^*, v^*) = (\frac{a}{5}, \frac{a^2}{25} + 1),
\end{equation}
derives as
\begin{equation}
\left. \boldsymbol{J}_0\right|_{u^*, v^*} =  
\frac{1}{a^2 + 25}
\begin{pmatrix}
\frac{3a^2 - 125}{\sigma} & -\frac{20a}{\sigma}\\
2a^2 b& - 5ab
\end{pmatrix}.
\end{equation}
First, note that the positive factor $1/(a^2 + 25)$ can be eliminated from all the following inequalities. Thus, to simplify the calculations, we do not include it in what follows.

Next, the rows and columns intervals write
\begin{equation}    
\begin{matrix}
\begin{pmatrix}
\frac{3a^2 - 125}{\sigma} & \qquad &  \qquad &  -\frac{20a}{\sigma}\\
\qquad &  \qquad &  \qquad & \qquad\\
2a^2 b&  \qquad &  \qquad &   - 5ab
\end{pmatrix} &
\begin{matrix}
\implies  [\frac{3a^2 -20a - 125}{\sigma}, \frac{3a^2 + 20a - 125}{\sigma}]\\
\quad & \quad\\
\implies  [- 5ab-2a^2 b, - 5ab+2a^2 b]
\end{matrix} \\
\begin{matrix}
\Downarrow & \Downarrow & \quad \\
[\frac{3a^2 - 125 -2\sigma a^2b}{\sigma}, \frac{3a^2 - 125 +2\sigma a^2b}{\sigma}] &
[\frac{- 5\sigma ab - 20a}{\sigma}, \frac{- 5\sigma ab + 20a}{\sigma}]
\end{matrix} & \quad \\
\end{matrix}. 
\end{equation}

For rows, we can see the center of the second row is always negative. As a result, we can have two conclusive cases:
\begin{enumerate}
	\item The center of the first row is on the positive side of the real line, while the other circle does not overlap the first one. These conditions can be written as 
	\begin{equation}
	\begin{cases}
	\frac{3a^2  - 125}{\sigma}>   0,\\
	\frac{3a^2 -20a - 125}{\sigma} >  -5ab + 2a^2 b ,\\  
	\end{cases},
	\end{equation}
	which simplify to two conditions:
	\begin{equation}
	a > \sqrt{\frac{125}{3}}, \qquad
	b < \frac{3 a^2 - 20 a - 125}{\sigma a (2 a - 5)}.
	\end{equation}
	Notice that the excluded region is $\sigma$-dependent.
	\item In the second case, the center of both circles is on the negative side of the real line, and $U_i \le 0$ for each. Hence, these conditions express in the following inequalities
	\begin{equation}
	\begin{cases}
	-5ab <   0,\\
	5ab \ge 2a^2b ,\\  
	\frac{3a^2  - 125}{\sigma} <   0,\\
	-\frac{(3a^2  - 125)}{\sigma} \ge \frac{20 a}{\sigma}  \\
	\end{cases}.
	\end{equation}
	The first inequality is always satisfied. And the remaining ones find an upper bound for $a$
	\begin{equation}
	\begin{cases}
	a \le \frac{5}{2} ,\\  
	a < \sqrt{\frac{125}{3}},\\
	3a^2 + 20 a -125 \le 0\\
	\end{cases}.
	\end{equation}
	Finally, the three inequalities reduce to one
	\begin{equation}
	0 < a < \frac{5}{2}.
	\end{equation}
\end{enumerate}
Both cases are depicted in Fig. \ref{fig:5}A and \ref{fig:5}B, respectively.

\begin{figure}
	\centering
	\begin{minipage}{0.45\textwidth}
		\centering
		\begin{tikzpicture}
		\draw[] (7, 3.5) node[above]{(A)};
		\draw[->] (8.4,0) -- (14.2,0) node[right] {$a$};
		\draw[-] (8.5,0) -- (8.5,0) node[left] {$\dots$};
		\draw[->] (7.9,-.5) -- (7.9,2.8) node[above] {$b$};
		\draw[domain=-0.5:2.5,smooth,variable=\b]
		plot ({8.5}, {\b}) node[right] {$a = \sqrt{\frac{125}{3}}$}; 
		
		\draw[domain=10.598:14,smooth,variable=\a]
		plot ({\a},{(3 * \a * \a - 20 * \a -125)/(.5*\a * (2 * \a - 5 ))}) node[right] {$\sigma=1$}; 
		
		\draw[domain=10.598:14,smooth,variable=\a]
		plot ({\a},{(3 * \a * \a - 20 * \a -125)/(.5*\a * (2 * \a - 5 ))}) node[right] {$\sigma=1$}; 
		
		\filldraw[pattern=crosshatch,domain=10.598:14,smooth,variable=\a] plot ({\a},{(3 * \a * \a - 20 * \a -125)/(.5*\a * (2 * \a - 5 ))}) -- (14, 0) -- (10.598, 0);
		
		\draw[domain=10.598:14.5,smooth,variable=\a]
		plot ({\a},{(3 * \a * \a - 20 * \a -125)/(1*\a * (2 * \a - 5 ))}) node[right] {$\sigma=2$}; 
		
		\end{tikzpicture}				
	\end{minipage}\hfill
	\begin{minipage}{0.45\textwidth}
		\centering
		\begin{tikzpicture}
		\draw[] (-.7, 3.5) node[above]{(B)};
		\draw[->] (-.2,0) -- (7.0,0) node[right] {$a$};
		\draw[->] (0,-.5) -- (0,2.8) node[above] {$b$};
		\draw[domain=-0.5:2.5,smooth,variable=\b]
		plot ({2.5}, {\b}) node[right] {$a = \frac{5}{2}$};
		\filldraw[pattern=crosshatch,domain=-0.5:2.5,smooth,variable=\b]
		plot ({2.5}, {\b}) -- (0, 2.5) -- (0, -.5);
		\end{tikzpicture}
	\end{minipage}\hspace*{\fill}
	\caption{\textbf{Lengyel-Epstein model}. (A) Row-induced parts of the parameter space (hatched area) that cannot produce Turing patterns. Note that it depends on $\sigma$. (B) The second result from the row-induced condition.}
	\label{fig:5}
\end{figure}
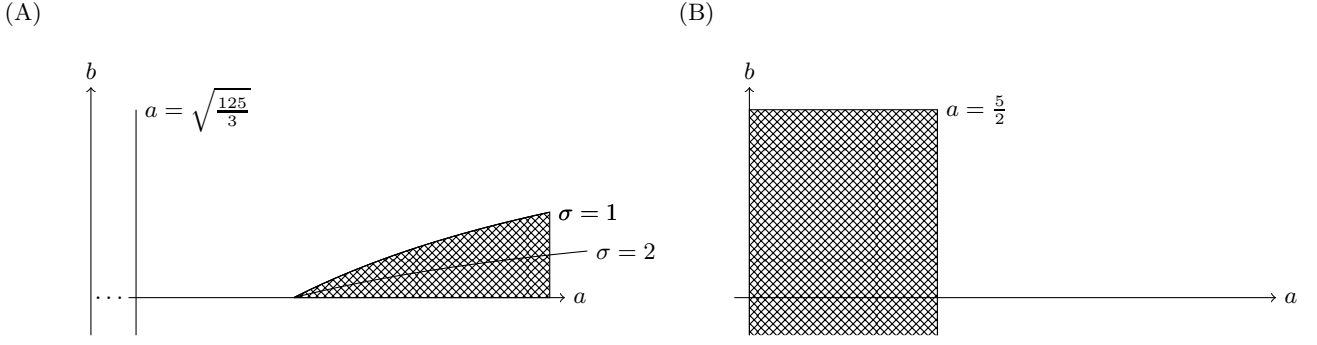

\begin{figure}
	\centering
	\begin{minipage}{0.45\textwidth}
		\centering
		\begin{tikzpicture}
		\draw[] (7, 3.5) node[above]{(A)};
		\draw[->] (8.4,0) -- (14.2,0) node[right] {$a$};
		\draw[-] (8.5,0) -- (8.5,0) node[left] {$\dots$};
		\draw[->] (7.9,-.5) -- (7.9,2.8) node[above] {$b$};
		\draw[domain=-0.5:2.5,smooth,variable=\b]
		plot ({8.5}, {\b}) node[right] {$a = \sqrt{\frac{125}{3}}$}; 
		
		\draw[domain=10.598:14,smooth,variable=\a]
		plot ({\a},{(3 * \a * \a - 20 * \a -125)/(.5*\a * (2 * \a - 5 ))}) node[right] {$\sigma=1$}; 
		
		\draw[domain=10.598:14,smooth,variable=\a]
		plot ({\a},{(3 * \a * \a - 20 * \a -125)/(.5*\a * (2 * \a - 5 ))}) node[right] {$\sigma=1$}; 
		
		\filldraw[pattern=crosshatch,domain=10.598:14,smooth,variable=\a] plot ({\a},{(3 * \a * \a - 20 * \a -125)/(.5*\a * (2 * \a - 5 ))}) -- (14, 0) -- (10.598, 0);
		
		\draw[domain=10.598:14.5,smooth,variable=\a]
		plot ({\a},{(3 * \a * \a - 20 * \a -125)/(1*\a * (2 * \a - 5 ))}) node[right] {$\sigma=2$}; 
		
		\end{tikzpicture}		
		
	\end{minipage}\hfill
	\begin{minipage}{0.45\textwidth}
		\centering
		\begin{tikzpicture}
		\draw[] (-.7, 3.5) node[above]{(B)};
		\draw[->] (-.2,0) -- (7.0,0) node[right] {$a$};
		\draw[->] (0,-.5) -- (0,2.8) node[left] {$b$};
		\draw[domain=0:6.5,smooth,variable=\a]
		plot ({\a}, {1}) node[right] {$b = \sqrt{\frac{4}{\sigma}}$}; 
		\draw[domain=-0.5:3.1,smooth,variable=\b]
		plot ({1.7}, {\b}) node[above] {$a = \sqrt{\frac{125}{3}}$}; 
		
		\draw[domain=1.5:6,smooth,variable=\a]
		plot ({\a},{(125 - 3 * \a * \a)/(10* 2* \a * \a )}) node[above] {$b = \frac{125 - 3a^2}{2\sigma a^2}$}; 
		\filldraw[pattern=crosshatch,domain=2.35:1.5,smooth,variable=\a] plot ({\a},{(125 - 3 * \a * \a)/(10* 2* \a * \a )}) -- (0, 2.6) -- (0, 1);
		
		\filldraw[pattern=crosshatch,domain=2.35:1.5,smooth,variable=\a] (1.7, 0) -- (1.7, 2.6) -- (0, 2.6) -- (0, 0);
		
		\end{tikzpicture}
		
	\end{minipage}\hspace*{\fill}
	\caption{\textbf{Lengyel-Epstein model}. (A) Column-induced parts of the parameter space (hatched area) that cannot produce Turing patterns. (B) The second result from the column-induced condition.}
	\label{fig:6}
\end{figure}
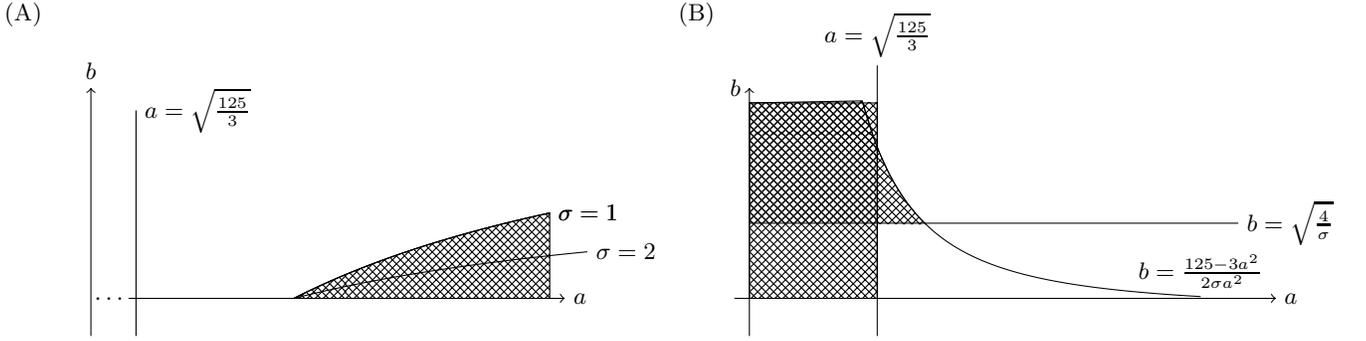
Again, the center of the second row is always negative for columns. Using the same argument and without reiterating it, we have the following cases:
\begin{enumerate}
	\item The center of the first column is on the positive side of the real line.
	\begin{equation}
	\begin{cases}
	\frac{3a^2  - 125}{\sigma}>   0,\\
	\frac{3a^2 - 2 \sigma a^2 b  - 125}{\sigma} >  \frac{-5 \sigma a b + 20 a}{\sigma} ,\\  
	\end{cases} \implies
	\begin{cases}
	a >   \sqrt{\frac{125}{3}},\\
	\frac{3a^2 - 20 a  - 125}{\sigma a (2a - 5)} >  b \\  
	\end{cases}.
	\end{equation}
	\item The center of both circles is on the negative side of the real line, and their radius is smaller than the distance between their center and the origin.
	\begin{equation}
	\begin{cases}
	-5ab <   0,\\
	5ab \ge \frac{20 a}{\sigma} ,\\  
	\frac{3a^2  - 125}{\sigma} <   0,\\
	-\frac{(3a^2  - 125)}{\sigma} \ge 2a^2b  \\
	\end{cases} \implies
	\begin{cases}		
	b \ge \frac{4}{\sigma} ,\\  
	a <  \sqrt{\frac{125}{3}},\\
	\frac{125 - 3a^2}{2\sigma a^2} \ge b  \\
	\end{cases}.
	\end{equation}
\end{enumerate}
Both cases are shown in Fig. \ref{fig:6}A and \ref{fig:6}B, respectively.

\end{document}